\documentclass{iopart}

\usepackage[colorlinks]{hyperref} %links for citations
\usepackage{graphicx}% Include figure files
\usepackage{mathptm}
\usepackage{iopams}

\newcommand{\ket}[1]{|#1\rangle}
\newcommand{\bra}[1]{\langle#1|}
\begin{document}

\title[Designing spin$-1$ lattice models using polar molecules]{Designing spin$-1$ lattice models using polar molecules}

\author{Gavin K. Brennen, Andrea Micheli, and Peter Zoller}
\address{Institute for Quantum Optics and Quantum Information of the Austrian Academy of Sciences,
6020, Innsbruck, Austria}\ead{Gavin.Brennen@uibk.ac.at}
\begin{abstract}
We describe how to design a large class of always on spin$-1$ interactions between polar molecules trapped in an optical lattice.  The spin degrees of freedom correspond to the hyperfine levels of a ro-vibrational ground state molecule.  Interactions are induced using a microwave field to mix ground states in one hyperfine manifold with the spin entangled dipole-dipole coupled excited states.  Using multiple fields anistropic models in one, two, or three dimensions, can be built with tunable spatial range.   An illustrative example in one dimension is the generalized Haldane model, which at a specific parameter has a gapped valence bond solid ground state.  The interaction strengths are large compared to decoherence rates and should allow for probing the rich phase structure of strongly correlated systems, including dimerized and gapped phases.  

\end{abstract}
%\maketitle
\maketitle

\section{Introduction}
Spin lattices are regular arrays of quantum mechanical spins with interactions involving small sets or neighborhoods of particles.  For a lattice of spin$-j$ particles with pairwise interactions, a generic
pairwise coupled spin Hamiltonian is written $H_{\rm spin}=\sum_{{\bf j},{\bf
    k}}\sum_{\alpha,\beta}\lambda^{\alpha}_{{\bf j}}X^{{\bf j},{\bf
    k}}_{\alpha,\beta} \lambda^{\beta}_{{\bf k}}$ where the collective
spatial indices ${\bf j},{\bf k}$ range over a D dimensional lattice
and the set of Hermitian operators $\{\lambda^{\alpha}\}_{\alpha=0}^{d^2-1}$
forms a representation of the algebra $\mathfrak{u}(2j+1)$.  The
interaction strength $X^{{\bf j},{\bf k}}_{\alpha,\beta}$ can depend
both on the magnitude and the direction of the difference vector ${\bf
  r}={\bf x}_{\bf j}-{\bf x}_{\bf k}$ between two lattice sites.   Such models capture the essential physics of more complicated systems and their relevance ranges from commercial materials fabrication to fundamental problems in physics.  Examples of the latter include studies of superconductivity and the origin of elementary particles and gauge fields \cite{Wen}.  

Because the Hilbert space dimension grows exponentially with the number of spins, predicting the kinematical and dynamical properties of spin systems is challenging and has been the focus of decades of research \cite{Auerbach:94}.  Some special classes of spin lattice models can be solved in closed form.  In those cases all properties of the associated quantum state and its dynamics are known.   In other cases, particularly one dimensional systems \cite{Vidal:03} and some finite correlated higher dimensional systems \cite{Cirac:06}, the ground states have an efficient approximate classical description which is suitable for estimating some relevant information about the quantum state such as correlation functions with bounded support.  In one dimensional quantum spin systems the dynamical evolution of a quantum state can also be efficiently described.  By efficient, it is meant that the propagator on an $n$ spin system can be computed with a number of resources that grows as a polynomial in $n$ provided the propagation time is no longer than polylogarithmic in $n$ \cite{Osborn:06}. 
Despite the notable progress in classical simulations, important properties of some spin lattices such as the behavior near a quantum phase transition which requires a long propagation time, or a function on the entire spectrum such as the free energy is required but an efficient classical algorithm is wanting.  In addition, there are some highly correlated ground states of spin lattice Hamiltonians that serve as resources for quantum information processing \cite{Cirac:04}.  For these reasons, it would be advantageous to be able to build a physical quantum mechanical simulation of the Hamiltonian.

One approach, the stroboscopic quantum simulator, works by mapping one quantum system to another using a sequence of local and coupling interactions \cite{Lloyd:96,Nielsen}.  Typically such simulations require a large overhead for spectroscopic approximation of one Hamiltonian, the target Hamiltonian, with another native to the simulator.   Making such a procedure fault tolerant necessitates an even larger overhead \cite{Brown:06}.  Another approach is to build the target Hamiltonian directly using ``always on" interactions.  This works by preparing a physical system including a set of particles endowed with a state space isomorphic to the target spins, and turning on a classical control field to build the target Hamiltonian.  The comparative advantage is a substantial reduction in control resources and time of the simulation.  However, given that it is an analog simulation, it is not amenable to error correction and hence is not fault tolerant.  Nevertheless, if the implemented interaction is close to the target interaction then many important properties of the system can be probed.
Examples of atom optical proposals of this kind include using tunneling dynamics of neutral atoms trapped in optical lattice to simulate spin exchange Hamiltonians \cite{Duan:03} and lattice gauge theories \cite{Buechler:04}.  

Prior work has proposed simulating integer spin models using neutral atoms trapped in optical lattices \cite{Cirac:04a},\cite{Demler:03}.  There the spin is encoded in electronic ground hyperfine levels and the effective interaction results from tunneling induced state dependent collisions with nearest neighbors.  A limitation of this mechanism is that the effective interaction strength $U$ is perturbative in the ratio of the tunneling rate $t$ to the collisional interaction $V$, scaling like $U\sim (t/V)^2 V$.  The ratio $t/V$ must be small to maintain the state in the projected subspace of one particle per lattice well and imposes a fundamental limitation to the effective spin interaction strength. The collisional interaction itself is typically $\lesssim 5$kHz limited to the available phase space density of two atoms occupying the motional ground state of a single lattice well.  It may be possible to build larger interactions using Feshbach resonances induced optically \cite{Julienne:06} or via shaped potentials \cite{Deutsch:03}, though it is not clear if such techniques will work for always on interactions.

Recently, we proposed an always on model to generate a large class of anisotropic spin$-1/2$ lattice models \cite{Micheli:06} using polar molecules trapped in an optical lattice.  Rapid progress has been made in the preparation and coherent control of cold dipolar molecules \cite{review} with an eye for electromagnetic trapping \cite{DeMille:05}.  The spectroscopy of the molecules is much richer than for single atoms allowing the simultaneous trapping on optical transition frequencies and coherent control at microwave frequencies with weak decoherence due to spontaneous emission.  These features make these systems experimentally relevant candidates for quantum simulators.  In our construction the spin is encoded in the valence electron of a $^{2}\Sigma$ polar molecule with zero nuclear spin prepared in the ro-vibrational ground state.  Spin-spin interactions can be induced by applying an external microwave field tuned near the transition between the ground and first rotational states to mix in dipole-dipole coupled states.  These rotationally excited states are superpositions of properly symmetrized spin entangled states, and by judicious choice of field strength, frequency and polarization, the coupling realizes  spin entangling interactions in the ground states.

Here we extend that work to build integer spin models.  This is a notable advance for two reasons.  First, integer spin lattice models have qualitatively different behavior from their half-integer spin counterparts.  In particular, antiferromagnetic Heisenberg interactions between one dimensional integer spin lattices exhibit a gap to excited states which is absent in the half-integer case.  This is the celebrated Haldane gap \cite{Haldane:88}.  Second, nature provides us with the state space isomorphic to interger spin by encoding in ground electronic hyperfine levels of molecules with half integer nuclear spin, of which there are many species.  Engineering spin$-j>1/2$ spin interactions is more challenging because the state space that that must be coherently controlled is significantly more complex.  A consequence of this structure is that special care must be taken to negate unwanted local interactions in the designed Hamiltonians.  Despite this, it is notable that closed form expressions for the effective interactions can be obtained in the asymptotic (large intermolecular separation) limit.  Furthermore, our analysis can be used as a guide for building more complex interactions with higher spin $F$ hyperfine states.
  
  The paper is organized as follows.  In section \ref{sec:spec} we introduce the state space of a single polar molecule with hyperfine structure in its low lying ro-vibrational states.  Depending on the total nuclear spin of the molecule, half spin or integer spin can be encoded into the hyperfine levels.  We focus on encoding in $F=1$ hyperfine states and derive the asymptotic dipole dipole coupling between a pair of $^{2}\Sigma$ molecules, one in the first excited rotational state one in the ground rotational state.   The effective spin interactions in the ground states are induced via microwave coupling to excited states and are derived in section \ref{sec:effHam}.  Because the interaction is additive, it is possible to  build design a variety of spin spin interactions using several fields tuned near excited state potentials.  We demonstrate an explicit construction of the general bilinear biquadratic isotropic Hamiltonian in one dimension.  In section \ref{meas} we describe how one might experimentally probe the many body ground state of a designed Hamiltonian.  Finally, we conclude with a summary and speculate about possible extensions of our construction.

     \begin{figure}
    \begin{center}
      \includegraphics[scale=0.6]{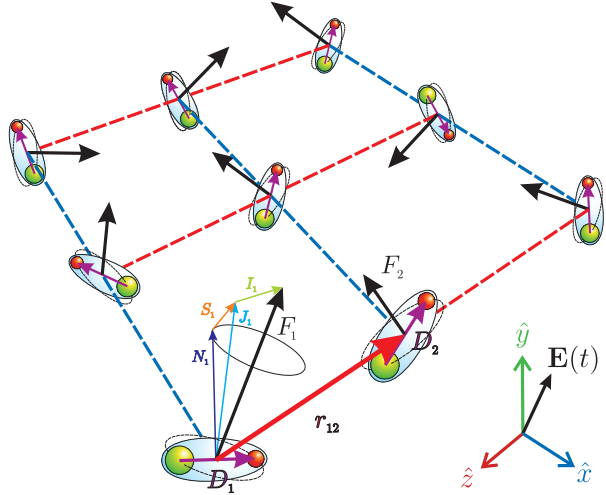}
      \caption{\label{fig:1}Setup depicting a two dimensional square lattice of polar molecules.   The lattice could correspond to an optical lattice constructed from two sets of orthogonal standing waves in the $\hat{x}$ and $\hat{y}$ directions and tight confinement in the $\hat{y}$ direction.  The lattice is shown illuminated by a spatially homogeneous linearly polarized microwave field ${\bf E}(t)$.  The spin-rotation and hyperfine couplings in each molecule are indicated.} 
    \end{center}
  \end{figure}

\section{The state space:  spectroscopy of dipolar molecules near ro-vibrational ground states}
\label{sec:spec}
The molecules we consider here are alkaline earth mono-halides with electronic ground states having $^{2}\Sigma$ symmetry.  These molecules, like alkali atoms, have a single valence electron allowing for optical trapping, e.g. in a far off resonant optical lattice.  Unlike the case with atoms, they also possess a finite dipole moment in their electronic ground states.  When prepared in a rotational eigenstate the spatially averaged dipole moment is zero, but it can be made non zero by mixing rotational states with opposite parity.    Because the rotational spacing as anharmonic it is possible to spectroscopically resolve the transition between ground and first excited rotational states and induce a finite dipole moment.  By including nuclear spin on one of the constituent atoms the hyperfine interaction defines a good quantum number $F$ which is the total electronic spin in the ground states.  For weak dipoles induced in states with $F=1$ the dipole-dipole interactions between pairs of molecules we thereby obtain  effective spin$-1$ Hamiltonians in the ground states.  It is a straightforward matter to adapt the following analysis to derive effective spin models with larger integer or half integer spins.
\subsection{Single body interactions}
The Hamiltonian describing the internal and external dynamics of a
pair of molecules trapped in wells of an optical lattice is 
$H=H_{\rm in}+H_{\rm ex}$.  The Hamiltonian describing the
external, or motional, degrees of freedom is $H_{\rm
ex}=\sum_{i=1}^2 {\bf P}_i^2/(2m)+ V_i({\bf x}_i-\bar{{\bf x}}_i)$,
where ${\bf P}_i$ is the momentum of molecule $i$ with mass $m$, and
the potential generated by the optical lattice
$V_i(\bf{x}-\bar{\bf{x}}_i)$ describes an external confinement of
molecule $i$ about a local minimum $\bar{\bf{x}}_i$ with $1$D rms
width $z_0$. We assume isotropic traps that are approximately
harmonic near the trap minimum with a vibrational spacing
$\hbar\omega_{\rm osc}$.  Furthermore, we assume that the molecules can be prepared in
the motional ground state of each local potential using dissipative
electromagnetic pumping \cite{review}.  It is convenient to
define the quantization axis $\hat{z}$ along the axis connecting the
two molecules, $\bar{\bf{x}}_2-\bar{\bf{x}}_1=\Delta z \hat{z}$ with
$\Delta z$ corresponding to a multiple of the lattice spacing.  We assume 
lattice trapping potentials that are strong relative to the energy scales of the 
dipole dipole coupled potentials.  This means that the mechanical forces
induced by coupling to the potentials do not significantly perturb the motional state of the 
ground states.  The optical lattice potential seen by the first excited rotational state
of the molecule is essentially identical to that seen by the ground rotational state.
This is in marked contrast to the situation of atoms coupled by optically induced dipole
dipole interactions where the excited states typically are antitrapped by the lattice
potential.   

\begin{figure}
\begin{center}
\includegraphics[scale=0.33]{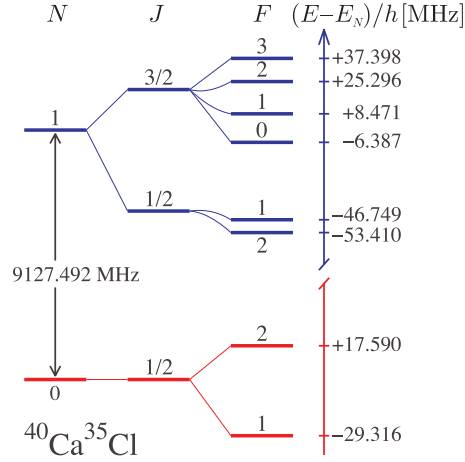}
\caption{\label{fig:2}
          Level-structure of a single $^{2}\Sigma$ polar molecule with nuclear spin $I=3/2$
            within the lowest two rotational manifolds
          $N=0,1$. The spectroscopic parameters for the example used in this work, 
          $^{40}$Ca$^{35}$Cl, are  $B=4563.746$MHz, $\gamma=42.208$MHz, $b=19.30134$MHz,
          $c=12.4554$MHz, $eqQ=1.00284$MHz, and $d=4.265$ Debye \cite{Moller:82}.  The bare 
          rotational energy is $E_N=BN(N+1)$. }
\end{center}
\end{figure}

The Hamiltonian for the internal
  degrees of freedom of two molecules is $H_{\rm int}=H_{\rm
    dd}+\sum_{j=1}^2 H^j_{\rm m}$, where $H_{\rm m}$ is the single
    body Hamiltonian $H_{\rm dd}$ is the dipole
    dipole coupling, described in detail below.  For molecules without nuclear spins,
    $H_{\rm m}=B{\bf N}^2+\gamma{\bf N}\cdot{\bf S}$ with $\bf N$ the
dimensionless orbital angular momentum of the nuclei, and $\bf S$ the
dimensionless electronic spin (assumed to be $S=1/2$ in the
following). Here $B$ denotes the rotational constant and $\gamma$ is
the spin-rotation coupling constant.  The coupled basis of a
single molecule $i$ corresponding to the eigenbasis of $H_{\rm m}^i$
is $\{\ket{N_i,S_i,J_i;M_{J_i}}\}$ where ${\bf J}_i={\bf N}_i+{\bf
  S}_i$ with eigenvalues $E(N=0,1/2,1/2)=0,E(1,1/2,1/2)=2B-\gamma$,
and $E(1,1/2,3/2)=2B+\gamma/2$.  
      Including effects due to nuclear
  spin {\bf I}, the single molecule Hamiltonian is (assuming for
  simplicity that only one of the two constituent atoms has a nuclear
  spin) \cite{Rad:64,Ryzlewicz:82}:
  \begin{equation}
  \begin{array}{lll}
    H^j_{\rm m}&=&B{\bf N}_j^2+\gamma {\bf N}_j\cdot {\bf S}_j+b{\bf I}_j\cdot {\bf S}_j+cI^z_jS^z_j\\
    & &+eQq[3I^{z 2}_j-I(I+1)]/4I(2I-1).
    \label{singlemolham}
    \end{array}
  \end{equation}
  The additional terms describe respectively: the Fermi contact,
  dipolar spin-spin, and electric quadrupole couplings.  The nuclear
  spin-rotation coupling is typically negligible for the $^{2}\Sigma$
  molecules of interest here.  A suitable basis set for the local interactions
  is: $\{\ket{N_j,S,J_j,I,F_j,M_{Fj}}\}$ where $H^j_{\rm m}$ 
  is diagonal in the quantum numbers $S,I,F,M_F$ but can couple states
  with different quantum numbers $N$ and $J$.  Expressions for these matrix
  element are given in \ref{AppA}.
We focus on the molecules where the rotational energy dominates over all other
internal energies in $H_{\rm m}$.  Therefore, we neglect
  couplings between different rotational states but include couplings
  $(J=N+1/2)\leftrightarrow (J=N-1/2)$. 
  Typically, the off diagonal matrix element is small compared to
  diagonal couplings meaning the electron spin is locked to the
  rotation.  The spectroscopy for an examplar molecule with $I=3/2$ is given in figure~\ref{fig:2}
  
%  The single molecule ground rotational states have the following couplings
%  \[
%  \begin{array}{lll}
%  &&\bra{N',1/2,J',3/2,1,M_F}H_{\rm m}\ket{0,1/2,1/2,3/2,1,M_F}=\\
%  &&\delta_{N',2}\delta_{J',3/2}(\sqrt{5} c/6-eQq/2\sqrt{5})\\
%  &&-\delta_{N',0}\delta_{J',3/2}5(b+c/3)/4.
% \end{array}
%  \]

We begin our derivation of the dipole-dipole potentials, and effective Hamiltonians
    to follow, with a diagonalization of the single body Hamiltonian $H_{\rm m}$ for 
    molecules with nuclear spin $I=3/2$.  The rotor piece of the single body Hamiltonian \ref{singlemolham} describes anharmonically spaced rotational states, so we restrict attention to the spectroscopically resolvable subspace $N=0,1$.  The ground states with quantum numbers
    $(N=0,J=1/2,F=1)$ have an energy 
    \begin{equation}
  E_{\rm gr}=-5 (b+c/3)/4
  \end{equation}
A rotationally excited molecule in $F=0$ has good quantum numbers $(N=1,J=1/2,F=0)$ and the energy is
  \begin{equation}
  E^{(0)}=2B+(2\gamma-5b-c-eQq)/4.
  \end{equation}
% For $^{40}{\rm Ca}{}^{35}{\rm Cl}$, $E^{(0)}/h=9121.11 {\rm MHz}$ and $E_{\rm gr}/h=-29.32 {\rm MHz}$. 
  Excited states with $F=1,2$ are linear combinations of states with $J=1/2$ and $J=3/2$.  For a given angular momentum projection $M_F$, the eigenstates of $H_{\rm m}$ are
\begin{equation}
 \begin{array}{lll}
 \ket{N=1,F,\pm,M_F}&=&\cos(\frac{\phi^{(F,\pm)}}{2})\ket{1,1/2,1/2,3/2,F,M_F}\\
 &+&\sin(\frac{\phi^{(F,\pm)}}{2})\ket{1,1/2,3/2,3/2,F,M_F},\\
 \end{array}
\end{equation}
 where the mixing angles are
 \begin{equation}
 \begin{array}{lll}
 \phi^{(1,-)}&=&\tan ^{-1}\left(\frac{4
   \sqrt{5} (10 b+5 c-3 eqQ)}{-90\gamma+80
   b-14 c+3 eqQ}\right),\\
    \phi^{(2,-)}&=&\tan ^{-1}\left(\frac{10 b+5 c+eqQ}{-30\gamma+6 c-3 eqQ}\right),
    \end{array}
   \end{equation}
   with $ \phi^{(F,+)}= \phi^{(F,-)}+\pi$.
   .  The eigenenergies are 
   \begin{equation}
   E^{(F,\pm)}=(v^{(F)}_1+v^{(F)}_2)/2\pm\sqrt{(v^{(F)}_1-v^{(F)}_2)^2/4+v^{(F)2}_3},
   \end{equation}
 where
 \[
 \begin{array}{lll}
 v^{(1)}_1&=&2B-(12\gamma-5b+5c)/12\\
 v^{(1)}_2&=&2B+(30\gamma-55b-11c-3eqQ)/60\\
 v^{(1)}_3&=&(10b+5c-3eqQ)/6\sqrt{5}\\
 v^{(2)}_1&=&2B-\gamma-(b-c)/4\\
 v^{(2)}_2&=&2B+\gamma/2-(5b+c-3eqQ)/20\\
 v^{(2)}_3&=&b+(5c+eqQ)/10.
 \end{array}
 \]
Excited states with $F=3$ are not dipole coupled to $F=1$ ground states and we do not include them here.  
%For 
 %  $^{40}{\rm Ca}{}^{35}{\rm Cl}$, 
   %$E_{\phi^{(1)}}/h=9080.75 {\rm MHz}, E_{\phi^{(1)}+\pi}/h=9135.97{\rm MHz}$ and 
%$\phi^{(1,-)}=-0.749215$.  

%For $^{40}{\rm Ca}{}^{35}{\rm Cl}$,
 %$E_{\phi^{(2)}}/h=9074.09 {\rm MHz}, E_{\phi^{(2)}+\pi}/h=9152.8{\rm MHz}$ and 
% the mixing angle is $\phi^{(2,-)}=-0.211354$.
 
%  The dipole matrix elements are
%\begin{equation}
%\begin{array}{lll}
%&&\bra{N=1,F=2,M_{F}+q;\ell}D^{\dagger}_q\ket{M_{F}}=(\delta_{q,-1}(\delta_{M_{F},-1}/\sqrt{6}+\delta_{M_{F},0}/2\sqrt{3}+\delta_{M_{F},1}/6)\\
%& &+\delta_{q,0}(-\delta_{M_{F},-1}/2\sqrt{3}+\delta_{M_{F},0}/3+\delta_{M_{F},1}/2\sqrt{3}\\
%& &+\delta_{q,1}(-\delta_{M_{F},-1}/6+\delta_{M_{F},0}/2\sqrt{3}+\delta_{M_{F},1}/\sqrt{6}))\\
%& &(\cos(\phi/2)-\sin(\phi/2)).
%\end{array}
%\end{equation}  
  
\subsection{Dipole-dipole interactions}
\label{dd}
      \begin{figure*}
    \begin{center}
      \includegraphics[width=0.99\textwidth]{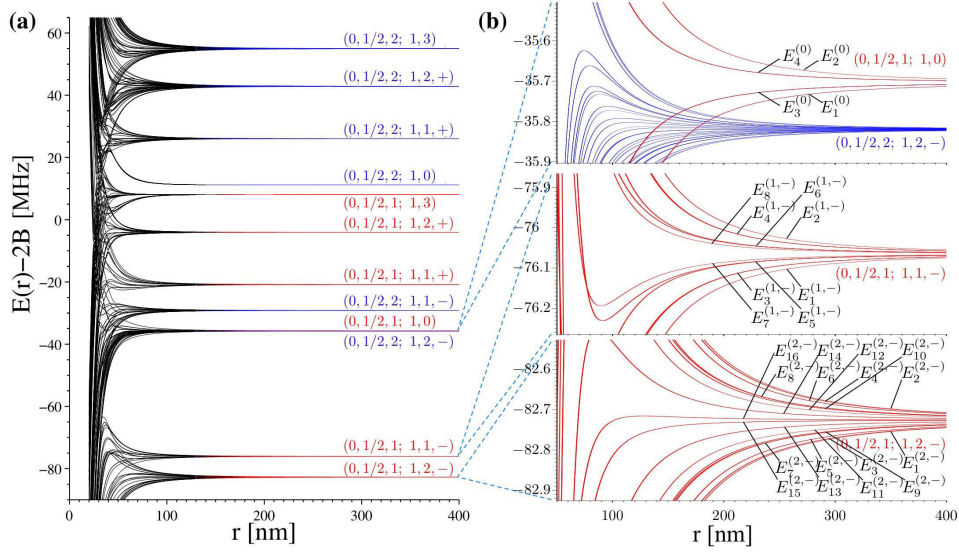}
      \caption{\label{fig:3}Dipole-dipole coupled
          excited rotation states for the subspace of one shared rotational quantum $N_1+N_2=1$ plotted as   
          a  function of intermolecular separation $r$.
           (a)  Excited states for $^{40}{\rm Ca}{}^{35}{\rm Cl}$ with 
          asymptotic quantum numbers $(N=0,J=1/2,F;N=1,F,\pm)$.  Red potential curves
          are dipole coupled to the $(0,1/2,1)$ ground states. (b) Magnification showing 
          sets of potential curves used in the text for building spin interactions.  This scale shows the 
          relevant range of pairwise separation at optical lattice spacings.}
    \end{center}
  \end{figure*}
  The near field dipole-dipole interaction between two molecules
separated by $r=|{\bf x}_1-{\bf x}_2|$ is
\begin{equation}
  H_{\rm dd}=
  \frac{d^2}{r^3}\Big(\sum_{q=-1}^1 (-1)^qD^{\dagger}_{1q}D_{2-q}-3D^{\dagger}_{10}D_{20}
  +h.c.\Big).
  \label{dd}
\end{equation}
The dipole operator coupling the ground and first rotational states of
molecule $i$ is ${\bf D}^{\dagger}_i=\sum_{q=-1}^1 \ket{N=1,q}_i
{_i}\bra{N=0,0}\hat{e}^*_q$, and $d$ is the dimensionful dipole
moment.
  The important consideration here is that since the
  $H_{\rm dd}$ acts to couple rotational degrees of
  freedom of two molecules, these interactions will be manifest at
  hyperfine energy scales due to the coupling of spin to rotation.
  Since $b,c$ are smaller than $\gamma$ we have a hope to obtain
  tunable spin interactions at intermolecular separations obtainable 
  with an optical lattice.
  
  The explicit dipole-dipole interaction has been solved before for the sharing
  of one optical photon between two alikali atoms with hyperfine structure \cite{Brennen}.
  After including nuclear spin, $H_{\rm dd}$ loses some
  of its symmetry with respect to operations over spin and spatial degrees of 
  freedom, however it still respects conservation of total
  projection quantum number along the intermolecular axis: i.e. $M_{F_{
    \rm tot}}=M_{F1}+M_{F2}$.  Additionally, $[H_{\rm dd},S_{1,2}]=0$ where
  $S_{1,2}$ is the exchange operator acting on the two molecules with
  eigenvalue $\sigma=\pm 1$.  In the subspace of one rotational quanta
  shared between the molecules, a good basis for $H_{\rm int}$ is
\[
    \begin{array}{lll}
      \ket{0,1/2,1/2,I,F_i,M_{Fi};1,1/2,J_i,I,F'_i,M_{F'i};\sigma}\equiv&& \\
      \frac{1}{\sqrt{2}} \Big(\ket{0,1/2,1/2,I,F_i,M_{Fi}}\otimes \ket{1,1/2,J'_i,I,F'_i,M_{F'i}}&&\\
      +\sigma \ket{1,1/2,J'_i,I,F'_i,M_{F'i}}\otimes\ket{0,1/2,1/2,I,F_i,M_{Fi}}\Big).
    \end{array}
\]
  Denoting the set of variable quantum numbers 
  \[{\mathcal
    S}=\{F_i,F'_i,F_j,F'_j,J'_i,J'_j,M_{Fi},M_{F'i},M_{Fj},M_{F'j}\},
    \]
  the dipole-dipole interaction in the subspace of one shared 
  quantum of rotation is then
  \begin{equation}
    \begin{array}{lll}
      H_{{\rm dd}}&=&\frac{d^2}{r^3}\sum_{\mathcal S}
     \sum_{\sigma=
    \pm 1}A({\mathcal S},I)\sigma\\
      & &\ket{0,1/2,1/2,I,F_j,M_{Fj};1,1/2,J'_j,I,F'_j,M_{F'j};\sigma}\\
      & &\bra{0,1/2,1/2,I,F_i,M_{Fi};1,1/2,J'_i,I,F'_i,M_{F'i};\sigma},
    \end{array}
  \end{equation}
  where
\[\fl
    \begin{array}{lll}
     A({\mathcal S},I)&=&6\delta_{M_{Fi}+M_{F'i},M_{Fj}+M_{F'j}}\sqrt{[F_j,J'_j,F'_j,J'_j]}\\
      & &(-1)^{F_i+F_j+1+2I+1+J'_j+J'_i+2M_{Fj}-2M_{F'i}}\left\{\begin{array}{ccc}J'_j & 1 & 1/2 \\F_i & I & F'_j\end{array}\right\}\\
      &&\left\{\begin{array}{ccc}1 & 1 & 0 \\ 1/2 & 1/2 & J'_j\end{array}\right\}\left\{\begin{array}{ccc}J'_i & 1 & 1/2 \\F_j & I & F'_I \end{array}\right\}\left\{\begin{array}{ccc}1 & 1 & 0 \\ 1/2 & 1/2 & J'_I\end{array}\right\}\\
      & & \Big[\sum_{q=-1}^1(-1)^q \bra{F'_j,-M_{F'j};1,q}F_i, -M_{Fi}\rangle\\
      &&\bra{F_j,-M_{Fj};1,-q}F'_i,-M_{F'i}\rangle-3 \bra{F'_j,-M_{F'j};1,0}F_i, -M_{Fi}\rangle\\
      &&\bra{F_j,-M_{Fj};1,0}F'_i,-M_{F'i}\rangle\Big].
    \end{array}
\]
  The intermolecular potentials are computed by diagonalizing $H_{\rm
    int}$ in blocks with good quantum numbers $M_{F_{
    \rm tot}}$ and $\sigma$.  In figure~\ref{fig:3} we
  plot the excited state potentials for $^{40}{\rm Ca}{}^{35}{\rm Cl}$ with $I=3/2$.  At the
  intermolecular separation $r\approx 200$nm we find appreciable
  mixing of asymptotic hyperfine states.  For trapping near this separation or smaller, the mixing of levels gives the possibility of tuning to different spin weighted states.  There is
  also the possibility that the microwave fields could be tuned such
  that more distant molecular pairs have small coupling due to off
  resonant cancellation effects from states with energies above and
  below the field frequency.  The asymptotic energy eigenvalues and 
 degeneracies of the $H_{\rm dd}$ in the subspace of one ground rotational 
 molecule with $F=1$ and one rotationally excited molecule are given in \ref{AppB}.
  
\section{Effective pairwise spin-one Hamiltonians}
\label{sec:effHam}
 By applying a microwave field tuned near resonant to the dipole-dipole coupled excited state potentials, it is possible to engineer effective spin-spin interactions in the ground spin states of a pair of molecules.  In the rotating wave approximation, the
field interaction with the two molecules at positions ${\bf x}_1,{\bf x}_2$ is 
\begin{equation}
H_{\rm mf}=-\sum_{i=1}^2 (\hbar \Omega
{\bf D}^{\dagger}_i\cdot {\bf e}_Fe^{i({\bf k}_F\cdot {\bf
    x}_i-\omega_F t)}/2+h.c.).
    \end{equation}
Here the electric field of amplitude $E_0$ is characterized by the Rabi frequency $|\Omega|=d|E_0|/\hbar$, polarization ${\bf
  e}_F=\alpha_{-}\hat{e}_{-1}+\alpha_{0}\hat{e}_{0}+\alpha_{+}\hat{e}_{1}$,
$(\hat{e}_0\equiv\hat{z})$, and frequency $\omega_F$.  For molecules spaced by optical wavelengths,
the dipoles are excited in phase and we can set ${\bf k}_F\cdot {\bf x}_i=0$.  For convenience we choose the intermolecular axis along $\hat{z}$.  A lattice of $n$ spins with spatial extent $\ell$ will similarly oscillate in phase provided $\ell |{\bf k}_F|\ll 1$, which suggests that superradiance could occur.  In that state the spontaneous emission rate from the excited rotational state is amplified by a factor of $n$.  To set the scale for these rates, note that blackbody scattering rates at room temperature for this system are on the order of $10^{-3}$ Hz compared to optical scattering rates induced by the confining optical lattice which are near the rate $1$ Hz \cite{DeMille:05}.  Smaller blackbody scattering rates are possible at a lower environmental temperature.  Throughout, we operate in the low saturation limit for $H_{\rm mf}$, such that the dominate source of decoherence is from the optical trapping field and superradiance does not impose a strong limitation on the lattice size.

When the saturation to the excited states is small, the effective Hamiltonian acting on the ground states is obtained in second order perturbation theory as
\begin{equation}
  H_{\rm eff}(r)=\sum_{i,f}\sum_{n}\frac{\bra{g_f}H_{\rm mf}\ket{\psi_n(r)}\bra{\psi_n(r)}H_{\rm mf}\ket{g_i}}{\hbar\omega_F-(E_n(r)-E_{\rm gr})}\ket{g_f}\bra{g_i},
\label{effHam1}
\end{equation}
where $\{{\ket{g_i},\ket{g_f}}\}$ are ground states with $N_1=N_2=0$ and energy $E_{\rm gr}$
and $\{\ket{\psi_n(r)}\}$ are excited eigenstates of $H_{\rm int}$
with $N_1+N_2=1$ and with excitation energies $\{E_n(r)\}$.
The reduced interaction in the subspace of the spin degrees of freedom
is then obtained by tracing over the relative spatial coordinates of the two molecules:  $H_{\rm spin}=\langle H_{\rm eff}(r)\rangle_{\rm rel}$.  In practice it is valid to model the interaction in the point dipole limit such that $H_{\rm spin}=H_{\rm eff}(\Delta z)$ provided the ratio of the spatial extent of each trapped molecule to the lattice spacing is small $(z_0/\Delta z\lesssim 0.1)$  (see Ref. \cite{Micheli:06} for details).

In the following derivation, we ignore off resonant dipole-dipole couplings to the ground states with $F=2$.  For this approximation to be valid, we demand that effective interaction $H_{\rm spin}$ be much smaller than the ground rotational state hyperfine splitting, or $||H_{\rm spin}||\ll 2(b+c/3)$, where we adopt the operator norm $||O ||\equiv\sup_{\ket{\psi}}||O-\mathrm{Tr}[O]{\bf 1}_d/d||/||\ket{\psi}||$, for $d$ the dimension of the Hilbert space that the operators act on ($d=9$ for pairwise operators).  If multiple microwave fields are used we further demand that Raman processes that might couple $F=1$ to $F=2$ rotational ground states are far off resonant.  These terms are negligible for the interactions considered here.
   
The effective spin interaction in the ground states as obtained from  \ref{effHam1} is:
 \begin{equation}
 H_{\rm eff}(r)=\frac{\hbar|\Omega|}{4}\sum_{k=0}^2\sum_{j,\pm} s^{(k,\pm)}_j(r) A^{(k)}_j({\bf e}_F),
 \label{effHam2}
 \end{equation}
where the saturation amplitudes are 
\[
\begin{array}{lll}
s^{(0)}_j(r)&=&\frac{\hbar|\Omega|}{(\hbar\omega_F-(E^{(0)}_j(r)-E_{\rm gr}))},\\
s_j^{(1,\pm)}(r)&=&\frac{\hbar|\Omega|(\cos(\phi^{(1,\pm)}/2)-\sin(\phi^{(1,\pm)}/2)\sqrt{5})^2}{36(\hbar\omega_F-(E^{(1,\pm)}_j(r)-E_{\rm gr}))},\\
s_j^{(2,\pm)}(r)&=&\frac{\hbar|\Omega|(1-\sin(\phi^{(2,\pm)}))}{(\hbar\omega_F-(E^{(2,\pm)}_j(r)-E_{\rm gr}))}.
\end{array}
\]
The polarization dependent coupling matrices $A^{(k)}_j$ expressed in the separable basis $\{\ket{g_{M_F}}\otimes \ket{g_{M_F}'}\}$ are given in \ref{AppB}. 
For a pair of molecules separated along a rotated axis $z'=D^{1\dagger}(\beta_1,\beta_2,\beta_3)\hat{e}_0$, where $D^{j}$ is the spin$-j$ Wigner rotation, the effective Hamiltonian is obtained using the frame rotated coupling matrices:  $A({\bf e}_F)\rightarrow D^1\otimes D^1A(D^{1\dagger}{\bf e}_F)D^{1\dagger}\otimes D^{1\dagger}$.  When multiple microwave fields are used at different frequencies and possibly different polarizations, the interaction is additive and complex spin interactions can be built as shown below. 

 Many lattice models involving physical spin$-j$ particles are described by Hamiltonians spanned by elements (or powers thereof) of an $\mathfrak{su}(2)$ subalgebra of the full $\mathfrak{u}(2j+1)$ operator algebra.  The elements of the spin$-1$ irrep of $\mathfrak{su}(2)$ are: 
\[\fl
  S^{x}=\frac{1}{\sqrt{2}}
\left(
\begin{array}{lll}
 0 &1 & 0 \\
 1 & 0 &
   1\\
 0 & 1 & 0
\end{array}
\right),\quad S^y=\frac{1}{\sqrt{2}}\left(
\begin{array}{lll}
 0 & -i & 0
   \\
 i & 0 &
   -i \\
 0 & i & 0
\end{array}
\right),\quad 
S^z=\left(
\begin{array}{lll}
 1 & 0 & 0 \\
 0 & 0 & 0 \\
 0 & 0 & -1
\end{array}
\right).
\]

In table~\ref{tab:1} we tabulate some spin models that can be built using multiple microwave configurations.  The quantization axis $\hat{z}$ is defined to be the intermolecular axis, assuming point like particles.  We assume that each field is tuned near resonant to one dipole-dipole coupled potential such that only saturation to that state need be considered.  In reality there will be residual off resonant couplings which will produce a Hamiltonian that deviates from the single state coupling result.  In building many of these spin patterns, such as for the two following examples, some sets of potentials must be coupled to by two independent fields.  This can be accomplished using fields tuned to both the upper and lower asymptotic manifolds.

%\begin{table}[!h]
\begin{table}
\caption{\label{tab:1}  Some spin patterns that result from \ref{effHam2}.  The field polarization is given with respect to the intermolecular axis $\hat{z}$ and the frequency of each field is chosen to be near resonant with one excited state potential.  The interaction is $H_{\rm eff}(r)=\hbar|\Omega|s(r)A/4$ with $A$ the two body spin pattern operator.  The saturation weights are chosen to describes the case where all constituent fields have the same Rabi frequency.  The same spin patterns can be realized with different field strengths by adjusting the detunings.}
%\caption{default}
\begin{center}
\begin{tabular}{|c|c|c|c|}
%\begin{ruledtabular}{ccc}
\hline\hline
field & near resonant saturation & polarization & spin pattern \\
\hline
$1$ & $-2s^{(1,\pm)}_{3}(r)= s(r)$ & $\hat{z}$&$S^{z}_1S^z_2$\\
 \hline
$2$ & $ -s^{(2,\pm)}_{7}(r)/6=s(r)$ & $\hat{z}$&\\
\hline
$ 3$ &$ \frac{s^{(2,\pm)}_{14}(r)(4\sqrt{3}-7)}{12 (5\sqrt{3}-9)}=s(r) $ & $\hat{z}$&\\
 \hline
 $1$ & $-s^{(0)}_{3}(r)/9= s(r)$ & $\hat{x}$&$S^{z}_1S^z_2$ \\
 \hline
$2$ & $ -2s^{(1,\pm)}_{3}(r)=s(r)$ & $\hat{z}$&\\
\hline
$ 3$ &$ \frac{s^{(2,\pm)}_{5}(r)(4\sqrt{3}+7)}{12 (5\sqrt{3}+9)}=s(r) $ & $\hat{z}$&\\
 \hline\hline
$1$& $2s^{(1,\pm)}_{3}(r)= s(r)$ & $\hat{z}$&$S^{z2}_1S^{z2}_2$ \\
 \hline
$2$ &$ 6s^{(1,\pm)}_{7}(r)=s(r)$ & $\hat{z}$&\\
 \hline
$1$& $s^{(0)}_{3}(r)/9= s(r)$ & $\hat{x}$&$S^{z2}_1S^{z2}_2$ \\
 \hline
$2$ &$2s^{(1,\pm)}_{3}(r)= s(r)$ & $\hat{z}$&\\
 \hline\hline
$1$ & $s_1^{(0)}(r)2/9= s(r)$ & $\hat{z}$&$({\bf 1}-S^{z 2})^{\otimes 2}$ \\
 \hline
  $1$ & $-s^{(0)}_{4}(r)2/9= s(r)$ & $\hat{y}$&\\
 \hline
  $2$ & $2s^{(1,\pm)}_{6}(r)= s(r)$ & $\hat{x}$&\\
\hline\hline
$1$ & $s^{(0)}_1(r)/6= s(r)$ & $\hat{z}$&$(S^{x}_1S^x_2+S^y_1S^y_2)$ \\
 \hline
$2$ & $ -s^{(0)}_3(r)/9= s(r)$ & $\hat{z}$&\\
\hline
$3$ & $ s^{(0)}_4(r)/9=  s(r) $ & $\hat{z}$&\\
\hline
$4$ & $ -s^{(2,\pm)}_1(r)/10.93725= s(r) $ & $\hat{z}$&\\
\hline
$5$ & $ s^{(2,\pm)}_{11}(r)/4.93725 = s(r)$ & $\hat{z}$&\\
 \hline\hline
$1$ &  $-2s^{(0)}_1(r)/3= s(r)$ & $\hat{z}$&$({\bf S}_1\cdot {\bf S}_2)^2$ \\
 \hline
$2$ & $ s^{(1,\pm)}_3(r)= s(r)$ & $\hat{z}$&\\
\hline
$3$ & $s^{(2,\pm)}_1(r)/10.93725= s(r) $ & $\hat{z}$&\\
\hline
$4$ & $ -s^{(2,\pm)}_{11}(r)/4.93725= s(r)$ & $\hat{z}$&\\
\hline
$5$ & $-\frac{(4\sqrt{3}-7)s^{(2,\pm)}_{14}(r)}{12 (5\sqrt{3}-9)}=s(r) $ & $\hat{z}$&\\
\hline\hline
$1$ &  $s^{(0)}_3(r)/36= s(r)$ & $\hat{x}$&$(S^x_1S^x_2-3S^z_1S^z_2)$ \\
\hline
$2$ &  $2s^{(0)}_4(r)/27= s(r)$ & $\hat{x}$& \\
 \hline
$3$ & $ -s^{(1,\pm)}_1(r)/35.32= s(r)$ & $\hat{x}$&\\
\hline
$4$ & $1.183 s^{(1,\pm)}_2(r)= s(r) $ & $\hat{x}$&\\
\hline
$5$ & $2s^{(1,\pm)}_6(r)/3= s(r) $ & $\hat{x}$&\\
\hline
$6$ & $-s^{(2,\pm)}_{4}(r)/4.604=s(r) $ & $\hat{x}$&\\
\hline
$7$ & $ s^{(2,\pm)}_{5}(r)/900.67= s(r)$ & $\hat{x}$&\\
\hline
$8$ &  $ s^{(2,\pm)}_{6}(r)/21.56= s(r)$ & $\hat{x}$& \\
 \hline
$9$ & $  -s^{(2,\pm)}_{8}(r)/27=s(r)$ & $\hat{x}$&\\
\hline
$10$ & $ -s^{(2,\pm)}_{9}(r)/29.54= s(r)$ & $\hat{x}$&\\
\hline
$11$ & $ -s^{(2,\pm)}_{11}(r)/16.34= s(r)$ & $\hat{x}$&\\
\hline
$12$ & $ -s^{(2,\pm)}_{13}(r)/42.59= s(r)$ & $\hat{x}$&\\
\hline
$13$ & $ s^{(2,\pm)}_{16}(r)/7.15= s(r)$ & $\hat{x}$&\\
\hline
\end{tabular}
\end{center}
\end{table}
\subsubsection{Example 1}
By combining three sets of spin operators in table~\ref{tab:1} we obtain the general bilinear, biquadratic, isotropic interaction between a pair $(j,k)$ of spins:
\begin{equation}
H^{j,k}_{\theta}=U(\cos{\theta}{\bf S}_j\cdot {\bf S}_k+\sin{\theta} ({\bf S}_j\cdot {\bf S}_k)^2),
\label{iso}
\end{equation} 
for $U>0$.  
This construction requires $10$ microwave fields to allow for tunable parameters $U$ and $\theta$ .  We demonstrate below using numerical optimization that the model can be approximated with acceptable accuracy using $4$ fields.  
\subsubsection{Example 2}
The last spin pattern in table~\ref{tab:1} is created using $13$ fields all polarized along the same direction.  While this is a rather complicated construction it demonstrate the possibility of anisotropic interactions in a plane.  For a pair of spins trapped in the $\hat{x}-\hat{y}$ plane with relative coordinate ${\bf r}=r (\cos\gamma \hat{y}+\sin\gamma \hat{z})$, and with all fields polarized along $\hat{z}$, the Hamlltonian is:  
\begin{equation}
H_{\rm spin}=U (S_1^zS_2^z-3S_1^{\gamma}S_2^{\gamma+\pi}),
\end{equation}
 where $S_i^{\gamma}=S^x\cos\gamma+S_i^y\sin\gamma$ and $U=\hbar|\Omega|\langle s(r)\rangle_{\rm rel}/4$. 

\subsection{A many-body result:  the Haldane model}
We wish to extend the two body effective Hamiltonian derived above to a many body system.  In doing so we again operate in the limit of low saturation to excited dipole-dipole coupled states and we treat the interaction in the ground rotational states as arising from the addition of the effective interaction $H_{\rm spin}$ between molecular pairs \cite{future}. 

A well studied spin$-1$ model in one dimension is generalized Haldane model which involves the nearest neighbor interaction $H_{\theta}$ on a chain of $N$ spins:
\begin{equation}
H_{\theta}=\sum_{j=1}^N H^{j,j+1}_{\theta}
\end{equation}
Ground states of this spin model have a rich phase diagram (see e.g. \cite{Schollwock:96, Schmitt:98}).  For $-\pi/4<\theta<\pi/4$, the system is in the Haldane phase characterized by exponentially decaying two body spin correlation functions and a finite energy gap to excited states that persists in  thermodynamic limit.  In that limit, the ground state is unique but for finite chains, there can be a degeneracy.  At the point $\tan{\theta}=1/3$, each summand in  \ref{iso} corresponds to a projector onto the spin 2 decomposition of $1\otimes 1=0\oplus1\oplus2$ and the $4$ degenerate ground states of the finite linear chain are gapped valence bond solid (VBS) states \cite{AKLT:87}.  VBS states have a practical use serving as a teleportation channel using local measurements only \cite{Milburn}.  For $-3\pi/4<\theta<-\pi/4$ the ground states are two fold degenerate dimerized singlets with finite gap to excited states.  At the point $\theta=-3\pi/4$, it has been conjectured \cite{Fath:93} that there is a continuous phase transition to a ferromagnetic phase which for the parameter range $\pi/2<\theta<5\pi/4$ has a ground state with ferromagnetic order and gapless excitations.  Finally, for $\pi/4<\theta<\pi/2$ the systems is in a trimerized phase.

\subsection{Realization}
\begin{figure}
    \begin{center}
      \includegraphics[width=\columnwidth]{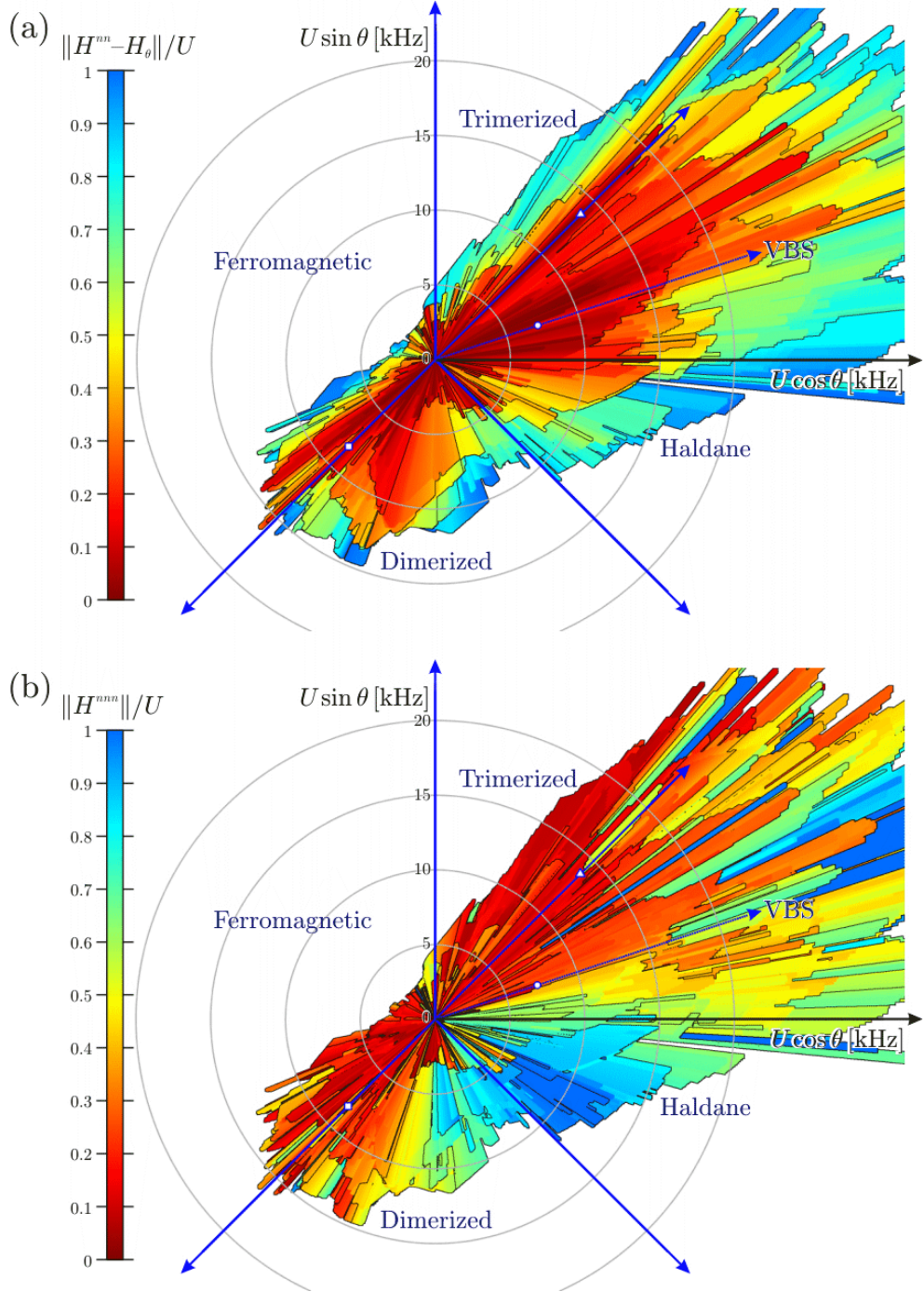}
      \caption{\label{fig:4}Phase diagram of ground states of $H_{\theta}$ for a chain of spin$-1$ particles.  The shaded areas correspond to regions can be prepared using engineered interactions between polar molecules trapped in an optical lattice.  The interactions $H^{nn}$, $H^{nnn}$ are the nearest neighbor and next nearest neighbor interactions obtained by numerical optimization over frequency, intensity, and polarization of four microwave fields.  The optimizations seek to closely approximate the desired two body nearest neighbor interaction $H_{\theta}$ (plot (a)) while minimizing next nearest (plot (b)) and longer range interactions.   The system parameters are set to those of $^{40}$Ca$^{35}$Cl molecules with a lattice spacing $\Delta z=200$nm.  In all the implementations, $\hat{z}$ polarized fields were optimal.   The strength of the interaction $U$ is constrained by demanding weak saturation to the excited dipole-dipole coupled states.   The $\bigtriangleup$, $\bigcirc$ and $\square$ indicate specific realizations with errors less than $0.05$. } 
    \end{center}
  \end{figure}

Consider a one dimensional lattice of trapped molecules illuminated by spatially homogeneous microwave coupling fields.  In practice the system could consist of molecules trapped in a three dimensional lattice with one dimension having a much smaller lattice spacing relative to other two such that the dominate pairwise interactions are along that dimension.  There is no a priori reason to expect that next nearest neighbor and longer range interactions can be ignored in 1D.  However, by applying multiple microwave fields, it is possible to substantially negate the effect of longer range interactions.  This occurs for two reasons:  first, the interaction in the ground states falls off not like $d^2/r^3$ but like $s(r)^2(\hbar\Delta-d^2/r^3))$, so that for longer range pairs, the interaction quickly saturates to a fixed value; second, using additional fields turned above or below resonance, the unwanted longer range interaction can be negated by the additivity of $H_{\rm eff}(r)$.  This mechanism also allows for the reduction in unwanted single body interactions.  We demonstrate this explicitly using numerical optimization to obtain a set of Hamiltonians close to $H_{\theta}$ (see figure~\ref{fig:4}).  The error of the implementations is measured by the operator distance of the implemented nearest neighbor Hamiltonian $H^{nn}$ to the closest isotropic Hamiltonian $H_{\theta}$.    The field parameters for three specific implementations are given in table~\ref{tab:2}.  The quality of the implementations do not change substantially using other molecular species, such as the $F=1$ hyperfine levels of CaF with nuclear spin $I=1/2$.

Notice that the high quality implementations tend to be clustered in the first and third quadrants of the plane in figure~\ref{fig:4}.  The reason for this bias is not completely understood, however, it could be partially attributed to the way we have chosen to parameterize $H_{\theta}$ i.e. using a decomposition over the non orthogonal operator basis ${\bf S}_1\cdot {\bf S}_2$ and $({\bf S}_1\cdot {\bf S}_2)^2$.  By comparison, we could construct an orthonomal operator basis $O_{1,2}$ such that $H=U'(\cos\theta 'O_1+\sin\theta 'O_2)$ where $O_1={\bf S}_1\cdot {\bf S}_2/2\sqrt{3}$, $O_2=({\bf S}_1\cdot {\bf S}_2+2 ({\bf S}_1\cdot {\bf S}_2)^2-8{\bf 1}_9/3)/2\sqrt{5}$, $U'\cos\theta '=\mbox{Tr}[O_1^{\dagger} H]$ and $U'\sin\theta '=\mbox{Tr}[O_2^{\dagger}H]$.  The probability distribution as a function of $\theta '$ for a random Hamiltonian with fixed strength $U'$ is $p(\theta ')d\theta '=d\theta '/2\pi$.  A fixed radius circle in the orthonormal basis is mapped to an ellipse in the non orthonormal basis which has an aspect ratio $[(10-2\sqrt{10})/(10+2\sqrt{10})]^{1/2}\approx 0.474$ and is rotated counter-clockwise in the plane by an angle $\tan^{-1}((1+\sqrt{10})/3)\approx 0.301 \pi$.  Hence, absent additional structure, we would expect to build Hamiltonians of fixed strength concentrated along an ellipse.  We hold to the original parameterization of  \ref{iso} to stay consistent with the literature.

   \begin{figure}
\begin{center}
\includegraphics[scale=0.4]{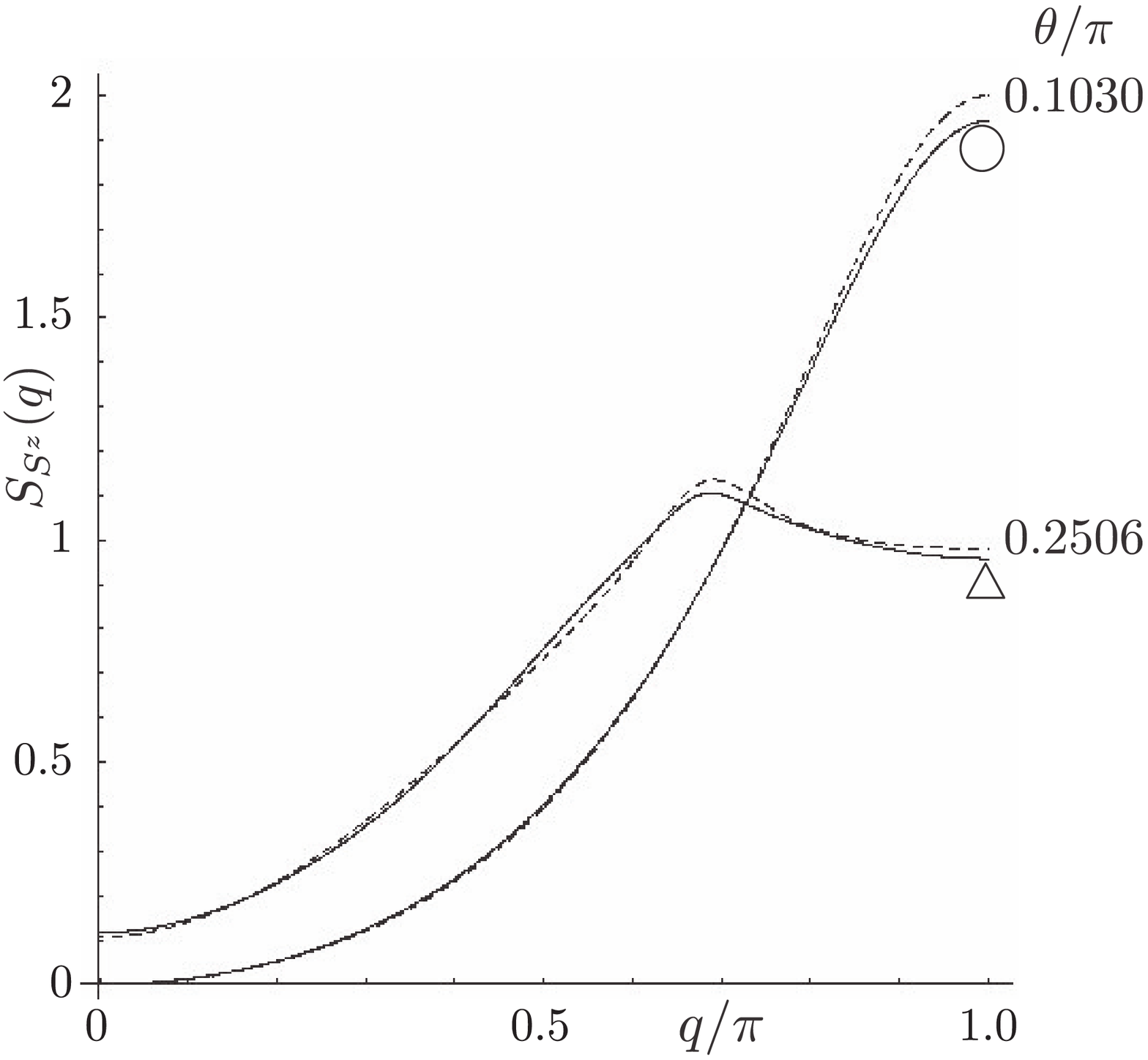}
\caption{\label{fig:5}
Spin structure factor for two implementations of nearly isotropic Hamiltonians on an infinite chain.  Solid lines correspond to the structure factors for the ground states of designed Hamiltonians with nearest neighbor interactions $H^{nn}_{\bigtriangleup}$, and $H^{nn}_{\bigcirc}$.   Dashed lines correspond to structure factors for the ground states of the closest isotropic Hamiltonian $H_{\theta}$ to each implementation.  The implementation $H_{\bigtriangleup}$ has a ground state near the Lai-Sutherland point \cite{Lai:74} where period three modes of spin order are predicted.  These are characterized by peaks in the spin structure factor at $q=\pm 2\pi/3$.}
\end{center}
\end{figure}
Important properties of spin lattice models can be obtained from static structure factors which are two body correlation functions.  The structure factors relative to the local operators $O_j$ for a chain of $N$ particles are defined:
\begin{equation}
S_O(q)=\frac{1}{N}\sum_{j,j'=1}^N e^{iq (j-j')}\langle O_jO_{j'}^{\dagger}\rangle,
\label{staticsf}
\end{equation}
where $q$ assumes the role of a quasimomentum for a one dimensional lattice of spins.
For the isotropic Hamiltonian $H_{\theta}$, relevant correlators involve the spin operator $O=S^z$ and center operators 
$O=e^{2i\pi S^Z/3}$ \cite{Schmitt:98}.  We have computed the correlations functions for the ground states of three specific implementations marked in figure~\ref{fig:4}.    The ground states themselves were computed using imaginary time evolution of the implemented Hamiltonians over a matrix product representation of an infinite chain of spin$-1$ particles \cite{Vidal:04}.  Results for the correlation function $S_{S^z}(q)$ for an infinite chain of polar molecules $(N\rightarrow \infty)$ are plotted in figure~\ref{fig:5}.  It is notable that even though the error of the implementations are non negligible, the spin correlation function is qualitatively quite close to that obtained with the closest isotropic model.  
\begin{table}[!h]
\caption{\label{tab:2}  Microwave field parameters for the realization of three implementations of the isotropic Hamiltonian $H_{\theta}$ in the ground $F=1$ manifold of $^{40}$Ca$^{35}$Cl.  All fields are polarized along $\hat{z}$, defined by the intermolecular axis. }
\begin{center}
\begin{tabular}{|c|c|c|c|}
%\begin{ruledtabular}{ccc}
\hline\hline
field & frequency $\omega_F$ [kHz] & amplitude $\Omega$ [kHz]  & spin pattern \\
\hline
$1$ & $E^{(2,-)}/h-120.619$ & $18.607$ & $H_{\bigcirc}$  \\
\hline
$2$ & $E^{(0)}/h-45.560$ & $19.862$ &   \\
\hline
$3$ & $E^{(2,-)}/h-188.368$ & $56.909$ &   \\
\hline
$4$ & $E^{(1,-)}/h-112.949$ & $61.667$ &   \\
\hline\hline
$1$ & $E^{(1,+)}h-38.456$ & $41.923$ & $H_{\bigtriangleup}$  \\
\hline
$2$ & $E^{(0)}/h-69.242$ & $13.064$ &   \\
\hline
$3$ & $E^{(1,+)}/h+111.868$ & $112.379$ &   \\
\hline
$4$ & $E^{(1,+)}/h-107.440$ & $84.709$ &   \\
\hline\hline
$1$ & $E^{(0)}/h-32.433$ & $17.240$ & $H_{\square}$  \\
\hline
$2$ & $E^{(0)}/h-90.400$ & $25.219$ &   \\
\hline
$3$ & $E^{(1,+)}/h-64.660$ & $19.929$ &   \\
\hline
$4$ & $E^{(0)}/h-29.087$ & $23.3787$ &   \\
\hline
\hline
\end{tabular}
\end{center}
\end{table}

\section{Spectroscopic measures of many body quantum phases}
\label{meas}

An essential component of always on quantum simulations will be measurement of properties of the many body state that is produced.  One such observable is the energy gap between ground and excited states.  If the system is assumed prepared in the ground state of a Hamiltonian $H$ and the excited states are coupled by single spin operators $S^{\alpha}_j$, then the gap can be probed with a locally coupling field as described in Ref. \cite{Micheli:06}.  Another set of observables are the spin structure factors \ref{staticsf}.  For our system, because of the small lattice spacing on the order of an optical wavelength, spin correlation functions may not be directly observable but can be inferred using time of flight measurements \cite{Grondalski:99,Altman:04}.  This can be accomplished by turning off the spin-spin interaction and the trapping fields which confine the molecules and allowing the particles to freely expand for a time $t$ after which they are measured by one or more detectors placed in a far field zone.  If the detectors can resolve different spin components then spatial-spin correlations are converted into momentum-spin correlations which can then be processed to infer relevant structure factors.
Spin resolving measurement are enabled by using a magnetic field gradient placed in the vicinity of a detection plane far from the source of molecules that spatially separates different hyperfine components.  This allows the measurement of first order correlation functions for a given spin component $\ket{\alpha}$, or second order correlations by observing coincidence counts for measuring a particle in state $\ket{\alpha}$ at detector $1$ and a particle in state $\ket{\beta}$ at detector $2$.  The latter describes a Hanbury-Brown Twiss type measurement and has been realized in the laboratory to measure spin independent two body correlation functions of atoms trapped in optical lattices \cite{Bloch:05}.  There the measurement corresponds to coincident count measurements on pixels of a CCD camera.  
%The secord order correlation function for joint detection of field intensities at momentum components ${\bf k},{\bf k}'$ is $G_{\alpha,\beta}({\bf r},{\bf r}')=\langle \Psi^{\dagger}_{\alpha}({\bf k})\Psi^{\dagger}_{\beta}({\bf k}')\Psi_{\beta}({\bf k}')\Psi^{\dagger}_{\alpha}({\bf k})\rangle$.  

For completeness, we describe in some detail here how to infer spin dependent correlation functions for spin$-1$ particles.  The system is a chain of molecules trapped in an optical lattice oriented along a space fixed $\hat{x}$ axis with lattice spacing $\lambda/2$ where $\lambda$ is the frequency of the trapping light.  In practice we may have an ensemble of identically prepared one dimensional systems trapped in a three dimensional optical lattice where the lattice spacing in the space fixed $\hat{y}-\hat{z}$ plane is much larger than $\lambda/2$ so that the chains do not interact with each other.  For simplicity, we assume the former case so that positions of the source particles and detectors are along the $\hat{x}$ axis only.  The latter can easily be treated by integrating the detected signal along a column. Consider a one dimensional lattice with $N$ molecules filling $N$ consecutive lattice sites with each molecule occupying the ground Wannier state centered at its local lattice well minimum.  When single molecules are released from the trap, they freely expand into a set of plane waves.
The atomic field at the detector plane is spanned by a set of $N$ discrete plane waves 
$\{k_j=j\pi\lambda/L^2\}_{j=-N/2}^{N/2}$ with mode spacing $\Delta k=\pi\lambda/L^2$, where $L=\sqrt{ht/M}$, and $M$ is the particle mass. 
The position space annihilation operator of a molecule with hyperfine spin component $\alpha$ at detector position $x_i$ is $b_{\alpha}(x_i)=\frac{1}{\sqrt{N}}\sum_j a_{\alpha}(X_j) e^{i k_jx_i}$.  Here $a_{\alpha}(X_i)$ is the annihilation operator for a molecule in state $\alpha$ in the ground Wannier state $w_j$ of the lattice well located at position $X_j=j\lambda/2$.    The second order correlation function describing coincidence counts at detector positions $x_1,x_2$ is then
\begin{equation}
\begin{array}{lll}
G^{(2)}_{\alpha,\beta}(x_1,x_2)&=&\langle b^{\dagger}_{\alpha}(x_1)b^{\dagger}_{\beta}(x_2)b_{\beta}(x_2)b_{\alpha}(x_1)\rangle\\
&=&\frac{V}{N^2}\displaystyle{\sum_{j,j',\ell,\ell '}}e^{i((k_j-k_{j'})x_1+(k_{\ell}-k_{\ell '})x_2)}\\
 &&\langle a^{\dagger}_{\alpha}(X_{j'})a^{\dagger}_{\beta}(X_{\ell '})a_{\beta}(X_{\ell})a_{\alpha}(X_j)\rangle\\
&=&\frac{V}{N^2}\displaystyle{\sum_{j,\ell}}[(1-\delta_{j,\ell})\langle n_{\alpha}(X_j) n_{\beta}(X_\ell)\rangle \\
&&+e^{iq_{1,2}(X_j-X_{\ell})}\langle a^{\dagger}_{\alpha}(X_{\ell})a^{\dagger}_{\beta}(X_j)a_{\beta}(X_{\ell})a_{\alpha}(X_j)\rangle],\\
\end{array}
\label{snd}
\end{equation}
where $V\simeq (L^2/4\pi z_0)^3$ is the volume of the freely evolved Wannier function in the detection plane.  In the third line we have used the fact that $\langle a^{\dagger}_{\alpha}(X_j)a_{\beta}(X_k)\rangle=\delta_{j,k}\langle a^{\dagger}_{\alpha}(X_j)a_{\beta}(X_j)\rangle$ for a unit filled lattice, and we have defined $q_{1,2}\equiv 2\pi(x_1-x_2)/L^2$.    For detectors that are not spin resolving we trace over spin to obtain
\begin{equation}
\begin{array}{lll}
G^{(2)}(x_1,x_2)&=&\sum_{\alpha,\beta}G^{(2)}_{\alpha,\beta}(x_1,x_2)\\
&=&\frac{V}{N^2}{\sum_{j,\ell}}[e^{iq_{1,2}(X_j-X_{\ell})}\langle S_{X_j,X_{\ell}}\rangle+(1-\delta_{j,\ell})]\\
%&=&\frac{V\eta}{N^2}{\sum_{j,\ell}}[e^{iq_{1,2}(X_j-X_{\ell})}\langle{\bf S}_j\cdot {\bf S}_{\ell}+({\bf S}_j\cdot {\bf S}_{\ell})^2\rangle\\
%&&+((1-\delta_{j,\ell})\eta-e^{iq_{1,2}(X_j-X_{\ell})})]\\
&=&V\Big(\eta N(N-1)/N^2-\frac{\sin^2(\pi q_{1,2} N\lambda/2)}{N^2\sin^2(\pi q_{1,2}\lambda/2)}\\
&&+\frac{\eta}{N^2}\displaystyle{\sum_{j,\ell}}e^{iq_{1,2}(X_j-X_{\ell})}\langle{\bf S}_j\cdot {\bf S}_{\ell}+({\bf S}_j\cdot {\bf S}_{\ell})^2\rangle\Big)
\end{array}
\end{equation}
where $\eta=1(-1)$ for bosons(fermions), $S_{X_j,X_{\ell}}$ is the swap operator
 between spins ${\bf S}_{j}$ and ${\bf S}_{{\ell}}$.
 
   In order to obtain the static structure factors \ref{staticsf} for operators diagonal in the $\hat{z}$ basis we need a measurement that will output expectation values $e^{iq_{1,2}(X_j-X_{\ell})}\langle a^{\dagger}_{\alpha}(X_{\ell})a^{\dagger}_{\beta}(X_j)a_{\beta}(X_{\ell})a_{\alpha}(X_{j})\rangle$.  The permutation symmetric components of these correlations can be obtained as per  \ref{snd} with $\alpha=\beta$.  We cannot obtain the other terms directly using measurements in the $\hat{z}$ basis.  However, these quantities can be obtained by measuring in a different local basis at each detector.  

First define the pairwise bases $\ket{x^{\alpha,\beta}_{\pm}}\equiv(\ket{\alpha}\pm\ket{\beta})/\sqrt{2}$,  and $\ket{y^{\alpha,\beta}_{\pm}}\equiv(\ket{\alpha}\pm i\ket{\beta})/\sqrt{2}$.  Consider the $q$ dependent terms in  \ref{snd} with $\alpha, \beta\in\{-1,1\}$ and $\alpha\neq \beta$.  If we measure in the $\ket{x^{\alpha,\beta}_\pm}_1$ and $\ket{y^{\alpha,\beta}_\pm}_1$ bases of the detector at position $x_1$ then we obtain the real and imaginary parts of the quantities: $e^{iq_{1,2}(X_j-X_{\ell})}\langle a^{\dagger}_{\alpha}(X_{\ell}) a^{\dagger}_{\beta}(X_j)a_{\beta}(X_{\ell})a_{\beta}(X_j)\rangle$.  Similarly, measuring in the $\ket{x^{\alpha,\beta}_{\pm}}_2$ and $\ket{y^{\alpha,\beta}_{\pm}}_2$ bases of the detector at position $x_2$ we obtain the real and imaginary parts of the quantities: $e^{iq_{1,2}(X_j-X_{\ell})}\langle a^{\dagger}_{\alpha}(X_{\ell})a^{\dagger}_{\alpha}(X_j)a_{\beta}(X_{\ell})a_{\alpha}(X_j)\rangle$.  Combining these results with those obtained with both detectors are measured in the $\ket{x^{\alpha,\beta}_{\pm}}_{1,2}$, $\ket{y^{\alpha,\beta}_{\pm}}_{1,2}$ and $\ket{z}_{1,2}$ bases, we obtain the desired quantities $e^{iq_{1,2}(X_j-X_{\ell})}\langle a^{\dagger}_{\alpha}(X_{\ell})a^{\dagger}_{\beta}(X_j)a_{\alpha}(X_{\ell})a_{\beta}(X_j)\rangle$.  Performing these measurements over all pairs 
$\alpha,\beta$ allows the computation of operators diagonal in the spin basis $\{\ket{m}\}$, and specifically the structure factors $S_{S^z}(q)$ and $S_{Z}(q)$.  Local measurement basis changes could be done in principle by using optical Raman pulses or microwave fields to selectively couple pairs of states in combination with magnetic field gradients.

\section{Conclusions}
Building on previous work on designing spin lattice Hamiltonians with spin$-1/2$ polar molecules we have shown how to build spin$-1$ models using the same mechanism.  The technique leverages off hyperfine structure in $^{2}\Sigma$ polar molecules with non zero nuclear spin, which is the most frequent situation found in nature.  Integer spin models are qualitatively different from half-integer spin models and accurate simulations of such models provide a wealth of new physics to explore, e.g. the nature of quantum phase transitions in the generalized Haldane model.  The results obtained here demonstrate that such models could be built using realistic lattice spacings $(\sim 250$nm) with interaction strengths on the order of $\simeq 10$kHz.  Decoherence is dominated by spontaneous emission due to optical excitation by the trapping lattice.  For the lattice spacing and localizations considered here the decoherence rate is $<1$ Hz \cite{DeMille:02} suggesting a quality factor of the implementation of $Q> 10^4$.  Stronger interactions are possible with smaller lattice spacings.  This could be achieved by eliminating the optical lattice altogether and opting for interaction induced confinement with self assembled dipolar crystals as proposed in Ref. \cite{Buchler:06}. 
 
While we have focused on optimizing one dimensional spin models, a more ambitious study would systematically attempt to design two dimensional spin$-1$ models.  Such models offer a host of new phenomena, among them ground states exhibiting topological order and emergent $U(1)$ gauge bosons \cite{Wen:03}.  We expect that an implementation using polar molecules would offer new possibilities for measurement and coherent control of such systems beyond that obtainable using condensed matter realizations.

\ack
We benefited from stimulating discussions with H.P. B\"uchler.  GKB thanks Jamie Williams for kindly providing Mathematica code to compute kinematics and dynamics of one dimensional spin chains.  This work was supported by the Austrian Science Foundation, the European Union under contract FP6-013501-OLAQUI, and the Institute for Quantum Information.

\appendix 
\section{Single molecule spectroscopy}
\label{AppA}
The matrix elements for the one body operator $H_{\rm m}$ are \cite{Rad:64,Ryzlewicz:82}:

  \begin{equation}
  \begin{array}{lll}
&&\bra{N',1/2,J',I,F,M_F}H_{\rm m}\ket{N,1/2,J,I,F,M_F}=\delta_{N,N'}\\
&&\Bigg[\delta_{J,J'}(BN(N+1)+\frac{\gamma}{2}
(J(J+1)-N(N+1)-3/4))\\
&&+(b+c/3)(-1)^{N+1/2+J+J'+I+F+1}\sqrt{[J,J',I]I(I+1)3/2}\\
&&\left\{\begin{array}{ccc}J' & 1/2 & N \\ 1/2 & J & 1 \end{array}\right\}
\left\{\begin{array}{ccc}F & J' & I \\ 1 & I & J \end{array}\right\}
\Bigg]+c\sqrt{5} (-1)^{J'+I+F+1}\\
&&\bra{N',0;2,0}N,0\rangle\left\{\begin{array}{ccc}F & J' & I \\ 1 & I & J \end{array}\right\}
\left\{\begin{array}{ccc}N & N' & 2 \\ 1/2 & 1/2 & 1 \\ J & J' & 1 \end{array}\right\}\\
&&\sqrt{[N',J,J',I]I(I+1)}
+eQq(-1)^{1/2+2J'+I+F-N}/4\\
&& \frac{\bra{N',0;2,0}N,0\rangle}{{\bra{I,-I;2,0}I,-I\rangle}}
\left\{\begin{array}{ccc}J' & N' & 1/2 \\ N & J & 2 \end{array}\right\}
\left\{\begin{array}{ccc}F & J' & I \\ 2 & I & J \end{array}\right\}\sqrt{[N',J,J',I]},
\end{array}
\end{equation}
where we adopt the notation $\{\cdots\}$ for the $6j$ and $9j$ symbols,  $\langle j_1,m_1;j_2,m_2|j_3,m_3 \rangle$ for the Clebsch Gordan coefficient, and $[A,B,\ldots,Z]=(2A+1)(2B+1)\cdots(2Z+1)$.

 Using the rules for angular momentum recoupling, the matrix elements of the dipole
 operator are:
\[\fl
 \begin{array}{lll}
 &&\bra{N,1/2,J,I,F,M_F}D^{\dagger}_q\ket{N',1/2,J',I,F',M_{F'}}=\sqrt{[F,J,J',N]}\\
 &&(-1)^{F'+N+1/2+I+J+J'+M_{F'}-M_F}\bra{F,-M_F;1,q} F',-M_{F'}\rangle\\
 & &\bra{N,0;1,0}N',0\rangle\left\{\begin{array}{ccc}J & 1 & J' \\F' & I & F \end{array}\right\}\left\{\begin{array}{ccc}N & 1 & N' \\J' & 1/2 & J \end{array}\right\}.
 \end{array}
\]

\section{Asymptotic dipole-dipole potentials}
\label{AppB}
\subsection{Coupling to $(N=0,F=1;N=1,F=0)$ asymptotic states}
\label{App1}
In the asymptotic regime where $||H_{\rm dd}(r)||\ll \gamma$, the dipole dipole interaction is
\begin{equation}
  H_{\rm dd}=\sum_{\sigma=\pm 1}\frac{\sigma d^2}{9r^3}(-2\ket{\lambda^0_{\sigma}}\bra{\phi^0_{\sigma}}
  +\ket{\lambda^1_{\sigma}}\bra{\lambda^1_{\sigma}}+\ket{\lambda^{-1}_{\sigma}}\bra{\lambda^{-1}_{\sigma}}),
\end{equation}
where $\ket{\lambda^{m}_{\sigma}}=\ket{0,1/2,1/2,3/2,1,m;1,1/2,3/2,3/2,0,0;\sigma}$.
The $4$ excited intermolecular potentials and corresponding degeneracies are 
\begin{equation}
\begin{array}{lll} 
 E^{(0)}_1(r)&=&E_{\rm gr}+E^{(0)}-2d^2/9r^3;\quad g_1=1\\
 E^{(0)}_2(r)&=&E_{\rm gr}+E^{(0)}+2d^2/9r^3;\quad g_2=1\\
 E^{(0)}_3(r)&=&E_{\rm gr}+E^{(0)}-d^2/9r^3;\quad g_3=2\\
 E^{(0)}_4(r)&=&E_{\rm gr}+E^{(0)}+d^2/9r^3;\quad g_4=2.\;
 \end{array}
\end{equation}

\subsection{Coupling to $(N=0,F=1;N=1,F=1,\pm)$ asymptotic states}
\label{App2}
The $8$ excited intermolecular potentials and respective degeneracies are

\begin{equation}
\begin{array}{lll}
E^{(1,\pm)}_1(r)&=&E_{\rm gr}+E^{(1,\pm)}-C^{(1,\pm)}(\sqrt{3}+1)d^2/36 r^3;\quad g_1=1\\
E^{(1,\pm)}_2(r)&=&E_{\rm gr}+E^{(1,\pm)}+C^{(1,\pm)}(\sqrt{3}+1)d^2/36 r^3;\quad g_2=1\\
E^{(1,\pm)}_3(r)&=&E_{\rm gr}+E^{(1,\pm)}-C^{(1,\pm)} 2 d^2/36 r^3;\quad g_3=3\\
E^{(1,\pm)}_4(r)&=&E_{\rm gr}+E^{(1,\pm)}+C^{(1,\pm)}2d^2/36 r^3;\quad g_4=3\\
E^{(1,\pm)}_5(r)&=&E_{\rm gr}+E^{(1,\pm)}-C^{(1,\pm)}d^2/36 r^3;\quad g_5=4\\
E^{(1,\pm)}_6(r)&=&E_{\rm gr}+E^{(1,\pm)}+C^{(1,\pm)}d^2/36 r^3;\quad g_6=4\\
E^{(1,\pm)}_7(r)&=&E_{\rm gr}+E^{(1,\pm)}-C^{(1,\pm)}(\sqrt{3}-1)d^2/36 r^3;\quad g_7=1\\
E^{(1,\pm)}_8(r)&=&E_{\rm gr}+E^{(1,\pm)}+C^{(1,\pm)}(\sqrt{3}-1)d^2/36 r^3;\quad g_8=1,\\
\end{array}
\end{equation}
where $C^{(1,\pm)}=(\cos(\phi^{(1,\pm)}/2)-\sin(\phi^{(1,\pm)}/2)\sqrt{5})^2$.

\subsection{Coupling to $(N=0,F=1;N=1,F=2,\pm)$ asymptotic states}
\label{App3}
The $16$ excited intermolecular potentials and corresponding degeneracies are 
\begin{equation}
\begin{array}{lll}
E^{(2,\pm)}_1(r)&=&E_{\rm gr}+E^{(2,\pm)}-C^{(2,\pm)} (7+\sqrt{7}) d^2/36r^3;\quad g_1=1\\
E^{(2,\pm)}_2(r)&=&E_{\rm gr}+E^{(2,\pm)}+C^{(2,\pm)}(7+\sqrt{7}) d^2/36r^3;\quad g_2=1\\
E^{(2,\pm)}_3(r)&=&E_{\rm gr}+E^{(2,\pm)}-C^{(2,\pm)}  0.257898 d^2/r^3;\quad g_3=2\\
E^{(2,\pm)}_4(r)&=&E_{\rm gr}+E^{(2,\pm)}+C^{(2,\pm)} 0.257898 d^2/r^3;\quad g_4=2\\
E^{(2,\pm)}_5(r)&=&E_{\rm gr}+E^{(2,\pm)}-C^{(2,\pm)}(\sqrt{3}+1)d^2/12r^3;\quad g_5=2\\
E^{(2,\pm)}_6(r)&=&E_{\rm gr}+E^{(2,\pm)}+C^{(2,\pm)}(\sqrt{3}+1)d^2/12r^3;\quad g_6=2\\
E^{(2,\pm)}_7(r)&=&E_{\rm gr}+E^{(2,\pm)}-C^{(2,\pm)}d^2/6r^3;\quad g_7=3\\
E^{(2,\pm)}_8(r)&=&E_{\rm gr}+E^{(2,\pm)}+C^{(2,\pm)} d^2/6r^3;\quad g_8=3\\
E^{(2,\pm)}_9(r)&=&E_{\rm gr}+E^{(2,\pm)}-C^{(2,\pm)} 0.156354 d^2/r^3;\quad g_9=2\\
E^{(2,\pm)}_{10}(r)&=&E_{\rm gr}+E^{(2,\pm)}+C^{(2,\pm)} 0.156354 d^2/r^3;\quad g_{10}=2\\
E^{(2,\pm)}_{11}(r)&=&E_{\rm gr}+E^{(2,\pm)}-C^{(2,\pm)} (7-\sqrt{7}) d^2/36r^3;\quad g_{11}=1\\
E^{(2,\pm)}_{12}(r)&=&E_{\rm gr}+E^{(2,\pm)}+C^{(2,\pm)} (7-\sqrt{7}) d^2/36r^3;\quad g_{12}=1\\
E^{(2,\pm)}_{13}(r)&=&E_{\rm gr}+E^{(2,\pm)}-C^{(2,\pm)} (\sqrt{3}-1) d^2/12r^3;\quad g_{13}=2\\
E^{(2,\pm)}_{14}(r)&=&E_{\rm gr}+E^{(2,\pm)}+C^{(2,\pm)} (\sqrt{3}-1) d^2/12r^3;\quad g_{14}=2\\
E^{(2,\pm)}_{15}(r)&=&E_{\rm gr}+E^{(2,\pm)}-C^{(2,\pm)}0.00956771 d^2/r^3;\quad g_{15}=2\\
E^{(2,\pm)}_{16}(r)&=&E_{\rm gr}+E^{(2,\pm)}+C^{(2,\pm)}0.00956771 d^2/r^3;\quad g_{16}=2.\\
\end{array}
\end{equation}
where $C^{(2,\pm)}=1-\sin(\phi^{(2,\pm)})$. 

\section{Catalogue of coupling matrices}
\label{AppC}

The effective spin Hamiltonians in the ground states are written in terms of coupling matrices $A^{(F)}_j$ weighting tensor coupling due to a field tuned near resonant with the excited state potential with energy $E^{(F,\pm)}_j(r)$.  These matrices are grouped according to the $(N-1,F)$ excited state manifold to which that the potentials asymptote.  The matrices are a function of the field polarization ${\bf e}_F=\alpha_-\hat{e}_{-1}+\alpha_0\hat{e}_0+\alpha_+\hat{e}_{1}$ and are expressed in the ground state product basis $\{\ket{-1}\ket{-1},\ket{-1}\ket{0}, \ket{-1}\ket{1},\ket{0}\ket{-1}, \ket{0}\ket{0},\ket{0},\ket{1}, \ket{1}\ket{-1},\ket{1}\ket{0}, \ket{1}\ket{1}\}$, where the quantization axis is the intermolecular axis.   

{\tiny{
 \[\fl
A^{(0)}_1=\left(
\begin{array}{lllllllll}
 0 & 0 & 0 & 0 & 0 & 0 & 0 & 0
   & 0 \\
 0 & \frac{|\alpha_+|^2}
   {18} & 0 & \frac{|\alpha_+|^2}
   {18} & -\frac{\alpha_0
   \alpha_+^{\ast}}
   {9} & \frac{\alpha_{-}
   \alpha_+^{\ast}}
   {18} & 0 & \frac{\alpha_{-}
   \alpha_+^{\ast}}
   {18} & 0 \\
 0 & 0 & 0 & 0 & 0 & 0 & 0 & 0
   & 0 \\
 0 & \frac{|\alpha_+|^2}
   {18} & 0 & \frac{|\alpha_+|^2}
   {18} & -\frac{\alpha_0
   \alpha_+^{\ast}}
   {9} & \frac{\alpha_{-}
   \alpha_+^{\ast}}
   {18} & 0 & \frac{\alpha_{-}
   \alpha_+^{\ast}}
   {18} & 0 \\
 0 & -\frac{\alpha_+
   \alpha_0^{\ast}}
   {9} & 0 & -\frac{\alpha_+
   \alpha_0^{\ast}}
   {9} & \frac{2 |\alpha_0|^2}
   {9} & -\frac{\alpha_{-}
   \alpha_0^{\ast}}
   {9} & 0 & -\frac{\alpha_{-}
   \alpha_0^{\ast}}
   {9} & 0 \\
 0 & \frac{\alpha_+
   \alpha_{-}^{\ast}
   }{18} & 0 & \frac{\alpha_+
   \alpha_{-}^{\ast}
   }{18} & -\frac{\alpha_0
   \alpha_{-}^{\ast}
   }{9} & \frac{|\alpha_{-}|^2
   }{18} & 0 &
   \frac{|\alpha_{-}|^2
   }{18} & 0 \\
 0 & 0 & 0 & 0 & 0 & 0 & 0 & 0
   & 0 \\
 0 & \frac{\alpha_+
   \alpha_{-}^{\ast}
   }{18} & 0 & \frac{\alpha_+
   \alpha_{-}^{\ast}
   }{18} & -\frac{\alpha_0
   \alpha_{-}^{\ast}
   }{9} & \frac{|\alpha_{-}|^2}{18} & 0 &
   \frac{|\alpha_{-}|^2
   }{18} & 0 \\
 0 & 0 & 0 & 0 & 0 & 0 & 0 & 0
   & 0
\end{array}
\right),
\]

\[\fl
A^{(0)}_2=\left(
\begin{array}{lllllllll}
 0 & 0 & 0 & 0 & 0 & 0 & 0 & 0
   & 0 \\
 0 & \frac{|\alpha_+|^2}
   {18} & 0 & -\frac{|\alpha_+|^2}
   {18} &0& -\frac{\alpha_{-}
   \alpha_+^{\ast}}
   {18} & 0 & \frac{\alpha_{-}
   \alpha_+^{\ast}}
   {18} & 0 \\
 0 & 0 & 0 & 0 & 0 & 0 & 0 & 0
   & 0 \\
 0 & -\frac{|\alpha_+|^2}
   {18} & 0 & \frac{|\alpha_+|^2}
   {18} & 0 & \frac{\alpha_{-}
   \alpha_+^{\ast}}
   {18} & 0 & -\frac{\alpha_{-}
   \alpha_+^{\ast}}
   {18} & 0\\
 0 & 0 & 0 & 0 & 0 & 0 & 0 & 0
   & 0 \\
 0 & -\frac{\alpha_+
   \alpha_{-}^{\ast}
   }{18} & 0 & \frac{\alpha_+
   \alpha_{-}^{\ast}
   }{18} &0& \frac{|\alpha_{-}|^2
   }{18} & 0 &
   -\frac{|\alpha_{-}|^2
   }{18} & 0 \\
 0 & 0 & 0 & 0 & 0 & 0 & 0 & 0
   & 0 \\
 0 & \frac{\alpha_+
   \alpha_{-}^{\ast}
   }{18} & 0 & -\frac{\alpha_+
   \alpha_{-}^{\ast}
   }{18} & 0 &- \frac{|\alpha_{-}|^2}{18} & 0 &
   \frac{|\alpha_{-}|^2
   }{18} & 0 \\
 0 & 0 & 0 & 0 & 0 & 0 & 0 & 0
   & 0
\end{array}
\right),
\]

\[\fl
A^{(0)}_3=\left(
\begin{array}{lllllllll}
 0 & 0 & 0 & 0 & 0 & 0 & 0 & 0
   & 0 \\
 0 & \frac{|\alpha_{0}|^2}
   {18} & -\frac{\alpha_{-}
   \alpha_{0}^{\ast}}
   {18} & -\frac{|\alpha_{0}|^2}
   {18} & 0 & 0 &
   \frac{\alpha_{-}
   \alpha_{0}^{\ast}}
   {18} & 0 & 0 \\
 0 & -\frac{\alpha_{0}
   \alpha_{-}^{\ast}
   }{18} & \frac{ |\alpha_+|^2+
   |\alpha_{-}|^2
  }{18} & \frac{\alpha_{0}
   \alpha_{-}^{\ast}
   }{18} & 0 &
   -\frac{\alpha_{0}
   \alpha_+^{\ast}}
   {18} &-\frac{ |\alpha_+|^2+
   |\alpha_{-}|^2
  }{18} & \frac{\alpha_{0}
   \alpha_+^{\ast}}
   {18} & 0 \\
 0 & -\frac{|\alpha_{0}|^2}
   {18} & \frac{\alpha_{-}
   \alpha_{0}^{\ast}}
   {18} & \frac{|\alpha_{0}|^2}
   {18} & 0 & 0 &
   -\frac{\alpha_{-}
   \alpha_{0}^{\ast}}
   {18} & 0 & 0 \\
 0 & 0 & 0 & 0 & 0 & 0 & 0 & 0
   & 0 \\
 0 & 0 & -\frac{\alpha_+
   \alpha_{0}^{\ast}}
   {18} & 0 & 0 &
   \frac{|\alpha_{0}|^2}
   {18} & \frac{\alpha_+
   \alpha_{0}^{\ast}}
   {18} & -\frac{|\alpha_{0}|^2}
   {18} & 0 \\
 0 & \frac{\alpha_{0}
   \alpha_{-}^{\ast}
   }{18} & -\frac{ |\alpha_+|^2+
   |\alpha_{-}|^2
  }{18}& -\frac{\alpha_{0}
   \alpha_{-}^{\ast}
   }{18} & 0 & \frac{\alpha_{0}
   \alpha_+^{\ast}}
   {18} & \frac{ |\alpha_+|^2+
   |\alpha_{-}|^2
  }{18}& -\frac{\alpha_{0}
   \alpha_+^{\ast}}
   {18} & 0 \\
 0 & 0 & \frac{\alpha_+
   \alpha_{0}^{\ast}}
   {18} & 0 & 0 &
   -\frac{|\alpha_{0}|^2}
   {18} & -\frac{\alpha_+
   \alpha_{0}^{\ast}}
   {18} & \frac{|\alpha_{0}|^2}
   {18} & 0 \\
 0 & 0 & 0 & 0 & 0 & 0 & 0 & 0
   & 0
\end{array}
\right),
\]

\[\fl
A^{(0)}_4=\left(
\begin{array}{lllllllll}
 \frac{2 |\alpha_+|^2}
   {9} & -\frac{\alpha_{0}
   \alpha_+^{\ast}}
   {9} & \frac{\alpha_{-}
   \alpha_+^{\ast}}
   {9} & -\frac{\alpha_{0}
   \alpha_+^{\ast}}
   {9} & 0 & 0 &
   \frac{\alpha_{-}
   \alpha_+^{\ast}}
   {9} & 0 & 0 \\
 -\frac{\alpha_+
   \alpha_{0}^{\ast}}
   {9} & \frac{|\alpha_{0}|^2}
   {18} & -\frac{\alpha_{-}
   \alpha_{0}^{\ast}}
   {18} & \frac{|\alpha_{0}|^2}
   {18} & 0 & 0 &
   -\frac{\alpha_{-}
   \alpha_{0}^{\ast}}
   {18} & 0 & 0 \\
 \frac{\alpha_+
   \alpha_{-}^{\ast}
   }{9} & -\frac{\alpha_{0}
   \alpha_{-}^{\ast}
   }{18} & \frac{ |\alpha_+|^2+
   |\alpha_{-}|^2
  }{18}
   & -\frac{\alpha_{0}
   \alpha_{-}^{\ast}
   }{18} & 0 &
   -\frac{\alpha_{0}
   \alpha_+^{\ast}}
   {18} & \frac{ |\alpha_+|^2+
   |\alpha_{-}|^2
  }{18} & -\frac{\alpha_{0}
   \alpha_+^{\ast}}
   {18} & \frac{\alpha_{-}
   \alpha_+^{\ast}}
   {9} \\
 -\frac{\alpha_+
   \alpha_{0}^{\ast}}
   {9} & \frac{|\alpha_{0}|^2}
   {18} & -\frac{\alpha_{-}
   \alpha_{0}^{\ast}}
   {18} & \frac{|\alpha_{0}|^2}
   {18} & 0 & 0 &
   -\frac{\alpha_{-}
   \alpha_{0}^{\ast}}
   {18} & 0 & 0 \\
 0 & 0 & 0 & 0 & 0 & 0 & 0 & 0
   & 0 \\
 0 & 0 & -\frac{\alpha_+
   \alpha_{0}^{\ast}}
   {18} & 0 & 0 &
   \frac{|\alpha_{0}|^2}
   {18} & -\frac{\alpha_+
   \alpha_{0}^{\ast}}
   {18} & \frac{|\alpha_{0}|^2}
   {18} & -\frac{\alpha_{-}
   \alpha_{0}^{\ast}}
   {9} \\
 \frac{\alpha_+
   \alpha_{-}^{\ast}
   }{9} & -\frac{\alpha_{0}
   \alpha_{-}^{\ast}
   }{18} & \frac{ |\alpha_+|^2+
   |\alpha_{-}|^2
  }{18}& -\frac{\alpha_{0}
   \alpha_{-}^{\ast}
   }{18} & 0 &
   -\frac{\alpha_{0}
   \alpha_+^{\ast}}
   {18} & \frac{ |\alpha_+|^2+
   |\alpha_{-}|^2
  }{18} & -\frac{\alpha_{0}
   \alpha_+^{\ast}}
   {18} & \frac{\alpha_{-}
   \alpha_+^{\ast}}
   {9} \\
 0 & 0 & -\frac{\alpha_+
   \alpha_{0}^{\ast}}
   {18} & 0 & 0 &
   \frac{|\alpha_{0}|^2}
   {18} & -\frac{\alpha_+
   \alpha_{0}^{\ast}}
   {18} & \frac{|\alpha_{0}|^2}
   {18} & -\frac{\alpha_{-}
   \alpha_{0}^{\ast}}
   {9} \\
 0 & 0 & \frac{\alpha_+
   \alpha_{-}^{\ast}
   }{9} & 0 & 0 &
   -\frac{\alpha_{0}
   \alpha_{-}^{\ast}
   }{9} & \frac{\alpha_+
   \alpha_{-}^{\ast}
   }{9} & -\frac{\alpha_{0}
   \alpha_{-}^{\ast}
   }{9} & \frac{2 |\alpha_{-}|^2
   }{9}
\end{array}
\right).
\]
}}
{\tiny{
\[\fl
A^{(1)}_1=
\left(
\begin{array}{lllllllll}
 0 & 0 & 0 & 0 & 0 & 0 & 0 & 0 & 0 \\
 0 & \frac{\left(-7+4 \sqrt{3}\right) |\alpha_+|^2}{4 \left(-3+\sqrt{3}\right)}
   & -\frac{\left(-2+\sqrt{3}\right) \alpha_0
   \alpha_+^{\ast}}{2 \left(-3+\sqrt{3}\right)}
   & \frac{\left(7-4 \sqrt{3}\right) |\alpha_+|^2}{4 \left(-3+\sqrt{3}\right)}
   & 0 & \frac{\left(-7+4 \sqrt{3}\right) \alpha_-
   \alpha_+^{\ast}}{4 \left(-3+\sqrt{3}\right)}
   & \frac{\left(-2+\sqrt{3}\right) \alpha_0
   \alpha_+^{\ast}}{2 \left(-3+\sqrt{3}\right)}
   & \frac{\left(7-4 \sqrt{3}\right) \alpha_-
   \alpha_+^{\ast}}{4 \left(-3+\sqrt{3}\right)}
   & 0 \\
 0 & -\frac{\left(-2+\sqrt{3}\right) \alpha_+
   \alpha_0^{\ast}}{2 \left(-3+\sqrt{3}\right)}
   & \frac{|\alpha_0|^2}{3-\sqrt{3}} &
   \frac{\left(-2+\sqrt{3}\right) \alpha_+
   \alpha_0^{\ast}}{2 \left(-3+\sqrt{3}\right)}
   & 0 & -\frac{\left(-2+\sqrt{3}\right) \alpha_-
   \alpha_0^{\ast}}{2 \left(-3+\sqrt{3}\right)}
   & \frac{|\alpha_0|^2}{-3+\sqrt{3}} &
   \frac{\left(-2+\sqrt{3}\right) \alpha_-
   \alpha_0^{\ast}}{2 \left(-3+\sqrt{3}\right)}
   & 0 \\
 0 & \frac{\left(7-4 \sqrt{3}\right) |\alpha_+|^2}{4 \left(-3+\sqrt{3}\right)}
   & \frac{\left(-2+\sqrt{3}\right) \alpha_0
   \alpha_+^{\ast}}{2 \left(-3+\sqrt{3}\right)}
   & \frac{\left(-7+4 \sqrt{3}\right) |\alpha_+|^2}{4 \left(-3+\sqrt{3}\right)}
   & 0 & \frac{\left(7-4 \sqrt{3}\right) \alpha_-
   \alpha_+^{\ast}}{4 \left(-3+\sqrt{3}\right)}
   & -\frac{\left(-2+\sqrt{3}\right) \alpha_0
   \alpha_+^{\ast}}{2 \left(-3+\sqrt{3}\right)}
   & \frac{\left(-7+4 \sqrt{3}\right) \alpha_-
   \alpha_+^{\ast}}{4 \left(-3+\sqrt{3}\right)}
   & 0 \\
 0 & 0 & 0 & 0 & 0 & 0 & 0 & 0 & 0 \\
 0 & \frac{\left(-7+4 \sqrt{3}\right) \alpha_+
   \alpha_-^{\ast}}{4
   \left(-3+\sqrt{3}\right)} &
   -\frac{\left(-2+\sqrt{3}\right) \alpha_0
   \alpha_-^{\ast}}{2
   \left(-3+\sqrt{3}\right)} & \frac{\left(7-4
   \sqrt{3}\right) \alpha_+
   \alpha_-^{\ast}}{4
   \left(-3+\sqrt{3}\right)} & 0 & \frac{\left(-7+4
   \sqrt{3}\right) |\alpha_-|^2}{4
   \left(-3+\sqrt{3}\right)} &
   \frac{\left(-2+\sqrt{3}\right) \alpha_0
   \alpha_-^{\ast}}{2
   \left(-3+\sqrt{3}\right)} & \frac{\left(7-4
   \sqrt{3}\right) |\alpha_-|^2}{4
   \left(-3+\sqrt{3}\right)} & 0 \\
 0 & \frac{\left(-2+\sqrt{3}\right) \alpha_+
   \alpha_0^{\ast}}{2 \left(-3+\sqrt{3}\right)}
   & \frac{|\alpha_0|^2}{-3+\sqrt{3}} &
   -\frac{\left(-2+\sqrt{3}\right) \alpha_+
   \alpha_0^{\ast}}{2 \left(-3+\sqrt{3}\right)}
   & 0 & \frac{\left(-2+\sqrt{3}\right) \alpha_-
   \alpha_0^{\ast}}{2 \left(-3+\sqrt{3}\right)}
   & \frac{|\alpha_0|^2}{3-\sqrt{3}} &
   -\frac{\left(-2+\sqrt{3}\right) \alpha_-
   \alpha_0^{\ast}}{2 \left(-3+\sqrt{3}\right)}
   & 0 \\
 0 & \frac{\left(7-4 \sqrt{3}\right) \alpha_+
   \alpha_-^{\ast}}{4
   \left(-3+\sqrt{3}\right)} &
   \frac{\left(-2+\sqrt{3}\right) \alpha_0
   \alpha_-^{\ast}}{2
   \left(-3+\sqrt{3}\right)} & \frac{\left(-7+4
   \sqrt{3}\right) \alpha_+
   \alpha_-^{\ast}}{4
   \left(-3+\sqrt{3}\right)} & 0 & \frac{\left(7-4
   \sqrt{3}\right) |\alpha_-|^2}{4
   \left(-3+\sqrt{3}\right)} &
   -\frac{\left(-2+\sqrt{3}\right) \alpha_0
   \alpha_-^{\ast}}{2
   \left(-3+\sqrt{3}\right)} & \frac{\left(-7+4
   \sqrt{3}\right) |\alpha_-|^2}{4
   \left(-3+\sqrt{3}\right)} & 0 \\
 0 & 0 & 0 & 0 & 0 & 0 & 0 & 0 & 0
\end{array}
\right),
\]

\[\fl
A^{(1)}_2=
\left(
\begin{array}{lllllllll}
 0 & 0 & 0 & 0 & 0 & 0 & 0 & 0 & 0 \\
 0 & \frac{3 |\alpha_+|^2}{12-4 \sqrt{3}} & 0 &
   \frac{3 |\alpha_+|^2}{12-4 \sqrt{3}} & 0 & \frac{3
   \alpha_- \alpha_+^{\ast}}{4 \left(-3+\sqrt{3}\right)} & 0 &
   \frac{3 \alpha_- \alpha_+^{\ast}}{4 \left(-3+\sqrt{3}\right)}
   & 0 \\
 0 & 0 & 0 & 0 & 0 & 0 & 0 & 0 & 0 \\
 0 & \frac{3 |\alpha_+|^2}{12-4 \sqrt{3}} & 0 &
   \frac{3 |\alpha_+|^2}{12-4 \sqrt{3}} & 0 & \frac{3
   \alpha_- \alpha_+^{\ast}}{4 \left(-3+\sqrt{3}\right)} & 0 &
   \frac{3 \alpha_- \alpha_+^{\ast}}{4 \left(-3+\sqrt{3}\right)}
   & 0 \\
 0 & 0 & 0 & 0 & 0 & 0 & 0 & 0 & 0 \\
 0 & \frac{3 \alpha_+ \alpha_-^{\ast}}{4
   \left(-3+\sqrt{3}\right)} & 0 & \frac{3 \alpha_+
   \alpha_-^{\ast}}{4 \left(-3+\sqrt{3}\right)} & 0 & \frac{3
   |\alpha_-|^2}{12-4 \sqrt{3}} & 0 & \frac{3
   |\alpha_-|^2}{12-4 \sqrt{3}} & 0 \\
 0 & 0 & 0 & 0 & 0 & 0 & 0 & 0 & 0 \\
 0 & \frac{3 \alpha_+ \alpha_-^{\ast}}{4
   \left(-3+\sqrt{3}\right)} & 0 & \frac{3 \alpha_+
   \alpha_-^{\ast}}{4 \left(-3+\sqrt{3}\right)} & 0 & \frac{3
   |\alpha_-|^2}{12-4 \sqrt{3}} & 0 & \frac{3
   |\alpha_-|^2}{12-4 \sqrt{3}} & 0 \\
 0 & 0 & 0 & 0 & 0 & 0 & 0 & 0 & 0
\end{array}
\right),
\]

\[\fl
A^{(1)}_3=
\left(
\begin{array}{lllllllll}
 2 |\alpha_0|^2 & -\alpha_-
   \alpha_0^{\ast} & 0 & -\alpha_- \alpha_0^{\ast} &
   0 & 0 & 0 & 0 & 0 \\
 -\alpha_0 \alpha_-^{\ast} & \frac{|\alpha_+|^2+2 |\alpha_-|^2}{4}
   & -\frac{\alpha_0 \alpha_+^{\ast}}{2} & \frac{ |\alpha_+|^2+2 |\alpha_-|^2}{4}
   & 0 & \frac{\alpha_-
   \alpha_+^{\ast}}{4} & -\frac{\alpha_0
   \alpha_+^{\ast}}{2} & \frac{\alpha_-
   \alpha_+^{\ast}}{4} & 0 \\
 0 & -\frac{\alpha_+ \alpha_0^{\ast}}{2} & |\alpha_0|^2 & -\frac{\alpha_+
   \alpha_0^{\ast}}{2} & 0 & -\frac{\alpha_-
   \alpha_0^{\ast}}{2} & |\alpha_0|^2 &
   -\frac{\alpha_- \alpha_0^{\ast}}{2} & 0 \\
 -\alpha_0 \alpha_-^{\ast} & \frac{|\alpha_+|^2+2 |\alpha_-|^2}{4} 
   & -\frac{\alpha_0 \alpha_+^{\ast}}{2} & \frac{ |\alpha_+|^2+2 |\alpha_-|^2}{4}
  & 0 & \frac{\alpha_-
   \alpha_+^{\ast}}{4} & -\frac{\alpha_0
   \alpha_+^{\ast}}{2} & \frac{\alpha_-
   \alpha_+^{\ast}}{4} & 0 \\
 0 & 0 & 0 & 0 & 0 & 0 & 0 & 0 & 0 \\
 0 & \frac{\alpha_+ \alpha_-^{\ast}}{4} & -\frac{\alpha_0
   \alpha_-^{\ast}}{2} & \frac{\alpha_+
   \alpha_-^{\ast}}{4} & 0 & \frac{2 |\alpha_+|^2+|\alpha_-|^2}{4}  &
   -\frac{\alpha_0 \alpha_-^{\ast}}{2} & \frac{2
   |\alpha_+|^2+|\alpha_-|^2}{4}  & -\alpha_0 \alpha_+^{\ast}
   \\
 0 & -\frac{\alpha_+ \alpha_0^{\ast}}{2} & |\alpha_0|^2 & -\frac{\alpha_+
   \alpha_0^{\ast}}{2} & 0 & -\frac{\alpha_-
   \alpha_0^{\ast}}{2} & |\alpha_0|^2 &
   -\frac{\alpha_- \alpha_0^{\ast}}{2} & 0 \\
 0 & \frac{\alpha_+ \alpha_-^{\ast}}{4} & -\frac{\alpha_0
   \alpha_-^{\ast}}{2} & \frac{\alpha_+
   \alpha_-^{\ast}}{4} & 0 & \frac{2 |\alpha_+|^2+|\alpha_-|^2}{4}  &
   -\frac{\alpha_0 \alpha_-^{\ast}}{2} & \frac{2
   |\alpha_+|^2+|\alpha_-|^2 }{4} & -\alpha_0 \alpha_+^{\ast}
   \\
 0 & 0 & 0 & 0 & 0 & -\alpha_+ \alpha_0^{\ast} & 0 & -\alpha_+
   \alpha_0^{\ast} & 2 |\alpha_0|^2
\end{array}
\right),
\]

\[\fl
A^{(1)}_4=
\left(
\begin{array}{lllllllll}
 0 & 0 & 0 & 0 & 0 & 0 & 0 & 0 & 0 \\
 0 & \frac{|\alpha_+|^2+2 |\alpha_-|^2}{4} & 0 & \frac{-|\alpha_+|^2-2 |\alpha_-|^2}{4}  
   & 0 & -\frac{\alpha_- \alpha_+^{\ast}}{4} & 0 &
   \frac{\alpha_- \alpha_+^{\ast}}{4} & 0 \\
 0 & 0 & 0 & 0 & 0 & 0 & 0 & 0 & 0 \\
 0 & \frac{-|\alpha_+|^2-2 |\alpha_-|^2}{4}  & 0 & \frac{|\alpha_+|^2+2 |\alpha_-|^2}{4}
   & 0 & \frac{\alpha_- \alpha_+^{\ast}}{4} & 0 &
   -\frac{\alpha_- \alpha_+^{\ast}}{4} & 0 \\
 0 & 0 & 0 & 0 & 0 & 0 & 0 & 0 & 0 \\
 0 & -\frac{\alpha_+ \alpha_-^{\ast}}{4} & 0 & \frac{\alpha_+
   \alpha_-^{\ast}}{4} & 0 & \frac{2 |\alpha_+|^2+|\alpha_-|^2}{4}  &
   0 & \frac{-2 |\alpha_+|^2-|\alpha_-|^2}{4}  & 0 \\
 0 & 0 & 0 & 0 & 0 & 0 & 0 & 0 & 0 \\
 0 & \frac{\alpha_+ \alpha_-^{\ast}}{4} & 0 & -\frac{\alpha_+
   \alpha_-^{\ast}}{4} & 0 & \frac{-2 |\alpha_+|^2-|\alpha_-|^2}{4}  &
   0 & \frac{ 2 |\alpha_+|^2+|\alpha_-|^2}{4} & 0 \\
 0 & 0 & 0 & 0 & 0 & 0 & 0 & 0 & 0
\end{array}
\right),
\]

\[\fl
A^{(1)}_5=
\left(
\begin{array}{lllllllll}
 0 & 0 & 0 & 0 & 0 & 0 & 0 & 0 & 0 \\
 0 & \frac{|\alpha_0|^2}{2} & 0 & -\frac{|\alpha_0|^2}{2} & 0 & 0 & 0 & 0 & 0 \\
 0 & 0 & \frac{|\alpha_+|^2+|\alpha_-|^2}{2}  & 0 & 0 & 0 & \frac{-|\alpha_+|^2-|\alpha_-|^2}{2}
  & 0 & 0 \\
 0 & -\frac{|\alpha_0|^2}{2} & 0 & \frac{|\alpha_0|^2}{2} & 0 & 0 & 0 & 0 & 0 \\
 0 & 0 & 0 & 0 & 0 & 0 & 0 & 0 & 0 \\
 0 & 0 & 0 & 0 & 0 & \frac{|\alpha_0|^2}{2} & 0 &
   -\frac{|\alpha_0|^2}{2} & 0 \\
 0 & 0 & \frac{-|\alpha_+|^2-|\alpha_-|^2}{2}  & 0 & 0 & 0 & \frac{|\alpha_+|^2+|\alpha_-|^2}{2}
   & 0 & 0 \\
 0 & 0 & 0 & 0 & 0 & -\frac{|\alpha_0|^2}{2} & 0 &
   \frac{|\alpha_0|^2}{2} & 0 \\
 0 & 0 & 0 & 0 & 0 & 0 & 0 & 0 & 0
\end{array}
\right),
\]

\[\fl
A^{(1)}_6=
\left(
\begin{array}{lllllllll}
 2 |\alpha_+|^2 & 0 & -\alpha_-
   \alpha_+^{\ast} & 0 & 0 & 0 & -\alpha_-
   \alpha_+^{\ast} & 0 & 0 \\
 0 & \frac{|\alpha_0|^2}{2} & 0 & \frac{|\alpha_0|^2}{2} & -\alpha_- \alpha_0^{\ast} &
   0 & 0 & 0 & 0 \\
 -\alpha_+ \alpha_-^{\ast} & 0 & \frac{|\alpha_+|^2+|\alpha_-|^2}{2}  &
   0 & 0 & 0 & \frac{|\alpha_+|^2+|\alpha_-|^2}{2}  &
   0 & -\alpha_- \alpha_+^{\ast} \\
 0 & \frac{|\alpha_0|^2}{2} & 0 & \frac{|\alpha_0|^2}{2} & -\alpha_- \alpha_0^{\ast} &
   0 & 0 & 0 & 0 \\
 0 & -\alpha_0 \alpha_-^{\ast} & 0 & -\alpha_0
   \alpha_-^{\ast} & 2 \left(|\alpha_+|^2+|\alpha_-|^2\right) &
   -\alpha_0 \alpha_+^{\ast} & 0 & -\alpha_0
   \alpha_+^{\ast} & 0 \\
 0 & 0 & 0 & 0 & -\alpha_+ \alpha_0^{\ast} & \frac{|\alpha_0|^2}{2} & 0 & \frac{|\alpha_0|^2}{2} & 0 \\
 -\alpha_+ \alpha_-^{\ast} & 0 & \frac{ |\alpha_+|^2+|\alpha_-|^2}{2} &
   0 & 0 & 0 & \frac{ |\alpha_+|^2+|\alpha_-|^2}{2} &
   0 & -\alpha_- \alpha_+^{\ast} \\
 0 & 0 & 0 & 0 & -\alpha_+ \alpha_0^{\ast} & \frac{|\alpha_0|^2}{2} & 0 & \frac{|\alpha_0|^2}{2} & 0 \\
 0 & 0 & -\alpha_+ \alpha_-^{\ast} & 0 & 0 & 0 & -\alpha_+
   \alpha_-^{\ast} & 0 & 2 |\alpha_-|^2
\end{array}
\right),
\]

\[\fl
A^{(1)}_7=
\left(
\begin{array}{lllllllll}
 0 & 0 & 0 & 0 & 0 & 0 & 0 & 0 & 0 \\
 0 & \frac{3 |\alpha_+|^2}{4 \left(3+\sqrt{3}\right)}
   & 0 & \frac{3 |\alpha_+|^2}{4
   \left(3+\sqrt{3}\right)} & 0 & -\frac{3 \alpha_-
   \alpha_+^{\ast}}{4 \left(3+\sqrt{3}\right)} & 0 & -\frac{3
   \alpha_- \alpha_+^{\ast}}{4 \left(3+\sqrt{3}\right)} & 0 \\
 0 & 0 & 0 & 0 & 0 & 0 & 0 & 0 & 0 \\
 0 & \frac{3 |\alpha_+|^2}{4 \left(3+\sqrt{3}\right)}
   & 0 & \frac{3 |\alpha_+|^2}{4
   \left(3+\sqrt{3}\right)} & 0 & -\frac{3 \alpha_-
   \alpha_+^{\ast}}{4 \left(3+\sqrt{3}\right)} & 0 & -\frac{3
   \alpha_- \alpha_+^{\ast}}{4 \left(3+\sqrt{3}\right)} & 0 \\
 0 & 0 & 0 & 0 & 0 & 0 & 0 & 0 & 0 \\
 0 & -\frac{3 \alpha_+ \alpha_-^{\ast}}{4
   \left(3+\sqrt{3}\right)} & 0 & -\frac{3 \alpha_+
   \alpha_-^{\ast}}{4 \left(3+\sqrt{3}\right)} & 0 & \frac{3
   |\alpha_-|^2}{4 \left(3+\sqrt{3}\right)} & 0 &
   \frac{3 |\alpha_-|^2}{4 \left(3+\sqrt{3}\right)}
   & 0 \\
 0 & 0 & 0 & 0 & 0 & 0 & 0 & 0 & 0 \\
 0 & -\frac{3 \alpha_+ \alpha_-^{\ast}}{4
   \left(3+\sqrt{3}\right)} & 0 & -\frac{3 \alpha_+
   \alpha_-^{\ast}}{4 \left(3+\sqrt{3}\right)} & 0 & \frac{3
   |\alpha_-|^2}{4 \left(3+\sqrt{3}\right)} & 0 &
   \frac{3 |\alpha_-|^2}{4 \left(3+\sqrt{3}\right)}
   & 0 \\
 0 & 0 & 0 & 0 & 0 & 0 & 0 & 0 & 0
\end{array}
\right),
\]

\[\fl
A^{(1)}_8=
\left(
\begin{array}{lllllllll}
 0 & 0 & 0 & 0 & 0 & 0 & 0 & 0 & 0 \\
 0 & \frac{\left(7+4 \sqrt{3}\right) |\alpha_+|^2}{4
   \left(3+\sqrt{3}\right)} & -\frac{\left(2+\sqrt{3}\right) \alpha_0
   \alpha_+^{\ast}}{2 \left(3+\sqrt{3}\right)} & -\frac{\left(7+4
   \sqrt{3}\right) |\alpha_+|^2}{4
   \left(3+\sqrt{3}\right)} & 0 & \frac{\left(7+4 \sqrt{3}\right) \alpha_-
   \alpha_+^{\ast}}{4 \left(3+\sqrt{3}\right)} &
   \frac{\left(2+\sqrt{3}\right) \alpha_0 \alpha_+^{\ast}}{2
   \left(3+\sqrt{3}\right)} & -\frac{\left(7+4 \sqrt{3}\right) \alpha_-
   \alpha_+^{\ast}}{4 \left(3+\sqrt{3}\right)} & 0 \\
 0 & -\frac{\left(2+\sqrt{3}\right) \alpha_+ \alpha_0^{\ast}}{2
   \left(3+\sqrt{3}\right)} & \frac{|\alpha_0|^2}{3+\sqrt{3}} & \frac{\left(2+\sqrt{3}\right)
   \alpha_+ \alpha_0^{\ast}}{2 \left(3+\sqrt{3}\right)} & 0 &
   -\frac{\left(2+\sqrt{3}\right) \alpha_- \alpha_0^{\ast}}{2
   \left(3+\sqrt{3}\right)} & -\frac{|\alpha_0|^2}{3+\sqrt{3}} & \frac{\left(2+\sqrt{3}\right)
   \alpha_- \alpha_0^{\ast}}{2 \left(3+\sqrt{3}\right)} & 0 \\
 0 & -\frac{\left(7+4 \sqrt{3}\right) |\alpha_+|^2}{4
   \left(3+\sqrt{3}\right)} & \frac{\left(2+\sqrt{3}\right) \alpha_0
   \alpha_+^{\ast}}{2 \left(3+\sqrt{3}\right)} & \frac{\left(7+4
   \sqrt{3}\right) |\alpha_+|^2}{4
   \left(3+\sqrt{3}\right)} & 0 & -\frac{\left(7+4 \sqrt{3}\right) \alpha_-
   \alpha_+^{\ast}}{4 \left(3+\sqrt{3}\right)} &
   -\frac{\left(2+\sqrt{3}\right) \alpha_0 \alpha_+^{\ast}}{2
   \left(3+\sqrt{3}\right)} & \frac{\left(7+4 \sqrt{3}\right) \alpha_-
   \alpha_+^{\ast}}{4 \left(3+\sqrt{3}\right)} & 0 \\
 0 & 0 & 0 & 0 & 0 & 0 & 0 & 0 & 0 \\
 0 & \frac{\left(7+4 \sqrt{3}\right) \alpha_+ \alpha_-^{\ast}}{4
   \left(3+\sqrt{3}\right)} & -\frac{\left(2+\sqrt{3}\right) \alpha_0
   \alpha_-^{\ast}}{2 \left(3+\sqrt{3}\right)} & -\frac{\left(7+4
   \sqrt{3}\right) \alpha_+ \alpha_-^{\ast}}{4
   \left(3+\sqrt{3}\right)} & 0 & \frac{\left(7+4 \sqrt{3}\right) |\alpha_-|^2}{4 \left(3+\sqrt{3}\right)} &
   \frac{\left(2+\sqrt{3}\right) \alpha_0 \alpha_-^{\ast}}{2
   \left(3+\sqrt{3}\right)} & -\frac{\left(7+4 \sqrt{3}\right) |\alpha_-|^2}{4 \left(3+\sqrt{3}\right)} & 0 \\
 0 & \frac{\left(2+\sqrt{3}\right) \alpha_+ \alpha_0^{\ast}}{2
   \left(3+\sqrt{3}\right)} & -\frac{|\alpha_0|^2}{3+\sqrt{3}} & -\frac{\left(2+\sqrt{3}\right)
   \alpha_+ \alpha_0^{\ast}}{2 \left(3+\sqrt{3}\right)} & 0 &
   \frac{\left(2+\sqrt{3}\right) \alpha_- \alpha_0^{\ast}}{2
   \left(3+\sqrt{3}\right)} & \frac{|\alpha_0|^2}{3+\sqrt{3}} & -\frac{\left(2+\sqrt{3}\right)
   \alpha_- \alpha_0^{\ast}}{2 \left(3+\sqrt{3}\right)} & 0 \\
 0 & -\frac{\left(7+4 \sqrt{3}\right) \alpha_+
   \alpha_-^{\ast}}{4 \left(3+\sqrt{3}\right)} &
   \frac{\left(2+\sqrt{3}\right) \alpha_0 \alpha_-^{\ast}}{2
   \left(3+\sqrt{3}\right)} & \frac{\left(7+4 \sqrt{3}\right) \alpha_+
   \alpha_-^{\ast}}{4 \left(3+\sqrt{3}\right)} & 0 &
   -\frac{\left(7+4 \sqrt{3}\right) |\alpha_-|^2}{4
   \left(3+\sqrt{3}\right)} & -\frac{\left(2+\sqrt{3}\right) \alpha_0
   \alpha_-^{\ast}}{2 \left(3+\sqrt{3}\right)} & \frac{\left(7+4
   \sqrt{3}\right) |\alpha_-|^2}{4
   \left(3+\sqrt{3}\right)} & 0 \\
 0 & 0 & 0 & 0 & 0 & 0 & 0 & 0 & 0
\end{array}
\right).
\]
}}

{\tiny{
\[\fl
A^{(2)}_1=10^{-3}\times\left(
\begin{array}{lllllllll}
 0 & 0 & 0 & 0 & 0 & 0 & 0 & 0 & 0 \\
 0 & 0.300215 \alpha_+ \alpha_+^{\ast} &
   -2.789446 \alpha_0 \alpha_+^{\ast} &
   0.300215 \alpha_+ \alpha_+^{\ast} &
   6.779751 \alpha_0 \alpha_+^{\ast} &
   0.300215 \alpha_- \alpha_+^{\ast} &
   -2.789446 \alpha_0 \alpha_+^{\ast} &
   0.300215 \alpha_- \alpha_+^{\ast} & 0 \\
 0 & -2.789446 \alpha_+ \alpha_0^{\ast} &
   25.918147 \alpha_0 \alpha_0^{\ast} &
   -2.789446 \alpha_+ \alpha_0^{\ast} &
   -62.994079 \alpha_0 \alpha_0^{\ast} &
   -2.789446 \alpha_- \alpha_0^{\ast} &
   25.918147 \alpha_0 \alpha_0^{\ast} &
   -2.789446 \alpha_- \alpha_0^{\ast} & 0 \\
 0 & 0.300215 \alpha_+ \alpha_+^{\ast} &
   -2.789446 \alpha_0 \alpha_+^{\ast} &
   0.300215 \alpha_+ \alpha_+^{\ast} &
   6.779751 \alpha_0 \alpha_+^{\ast} &
   0.300215 \alpha_- \alpha_+^{\ast} &
   -2.789446 \alpha_0 \alpha_+^{\ast} &
   0.300215 \alpha_- \alpha_+^{\ast} & 0 \\
 0 & 6.779751 \alpha_+ \alpha_0^{\ast} &
   -62.994079 \alpha_0 \alpha_0^{\ast} &
   6.779751 \alpha_+ \alpha_0^{\ast} &
   153.107164 \alpha_0 \alpha_0^{\ast} &
   6.779751 \alpha_- \alpha_0^{\ast} &
   -62.994079 \alpha_0 \alpha_0^{\ast} &
   6.779751 \alpha_- \alpha_0^{\ast} & 0 \\
 0 & 0.300215 \alpha_+ \alpha_-^{\ast} &
   -2.789446 \alpha_0 \alpha_-^{\ast} &
   0.300215 \alpha_+ \alpha_-^{\ast} &
   6.779751 \alpha_0 \alpha_-^{\ast} &
   0.300215 \alpha_- \alpha_-^{\ast} &
   -2.789446 \alpha_0 \alpha_-^{\ast} &
   0.300215 \alpha_- \alpha_-^{\ast} & 0 \\
 0 & -2.789446 \alpha_+ \alpha_0^{\ast} &
   25.918147 \alpha_0 \alpha_0^{\ast} &
   -2.789446 \alpha_+ \alpha_0^{\ast} &
   -62.994079 \alpha_0 \alpha_0^{\ast} &
   -2.789446 \alpha_- \alpha_0^{\ast} &
   25.918147 \alpha_0 \alpha_0^{\ast} &
   -2.789446 \alpha_- \alpha_0^{\ast} & 0 \\
 0 & 0.300215 \alpha_+ \alpha_-^{\ast} &
   -2.789446 \alpha_0 \alpha_-^{\ast} &
   0.300215 \alpha_+ \alpha_-^{\ast} &
   6.779751 \alpha_0 \alpha_-^{\ast} &
   0.300215 \alpha_- \alpha_-^{\ast} &
   -2.789446 \alpha_0 \alpha_-^{\ast} &
   0.300215 \alpha_- \alpha_-^{\ast} & 0 \\
 0 & 0 & 0 & 0 & 0 & 0 & 0 & 0 & 0
\end{array}
\right)
\]

\[\fl
A^{(2)}_2=
0.031797\times\left(
\begin{array}{lllllllll}
 0 & 0 & 0 & 0 & 0 & 0 & 0 & 0
   & 0 \\
 0 &
   |\alpha_+|^2
   & 0 &
   - |\alpha_+|^2
   & 0 &
   -\alpha_-
   \alpha_+^{\ast}
   & 0 &
   \alpha_-
   \alpha_+^{\ast}
   & 0 \\
 0 & 0 & 0 & 0 & 0 & 0 & 0 & 0
   & 0 \\
 0 &
   -|\alpha_+|^2
   & 0 &
   |\alpha_+|^2
   & 0 &
   \alpha_-
   \alpha_+^{\ast}
   & 0 &
   -\alpha_-
   \alpha_+^{\ast}
   & 0 \\
 0 & 0 & 0 & 0 & 0 & 0 & 0 & 0
   & 0 \\
 0 &
   -\alpha_+
   \alpha_-^{\ast}
   & 0 &
   \alpha_+
   \alpha_-^{\ast}
   & 0 &
   |\alpha_-|^2
   & 0 &
   -|\alpha_-|^2
   & 0 \\
 0 & 0 & 0 & 0 & 0 & 0 & 0 & 0
   & 0 \\
 0 &
   \alpha_+
   \alpha_-^{\ast}
   & 0 &
   -\alpha_+
   \alpha_-^{\ast}
   & 0 &
   -|\alpha_-|^2
   & 0 &
  |\alpha_-|^2
   & 0 \\
 0 & 0 & 0 & 0 & 0 & 0 & 0 & 0
   & 0
\end{array}
\right),
\]

\[\fl
A^{(2)}_3=
\left(
\begin{array}{lllllllll}
 0 & 0 & 0 & 0 & 0 & 0 & 0 & 0
   & 0 \\
 0 &
   0.092902 |\alpha_0|^2
   &
   0.008406\alpha_-
   \alpha_0^{\ast}
   &
   -0.092902 |\alpha_0|^2
   & 0 & 0 &
   -0.008406
   \alpha_-
   \alpha_0^{\ast}
   & 0 & 0 \\
 0 &
   0.008406\alpha_0
   \alpha_-^{\ast}
   &
   0.000761\left(|\alpha_+|^2+
   |\alpha_-|^2
   \right) &
   -0.008406
   \alpha_0
   \alpha_-^{\ast}
   & 0 &
   0.008406\alpha_0
   \alpha_+^{\ast}
   &
   -0.000761   \left(|\alpha_+|^2+
   |\alpha_-|^2
   \right) &
   -0.008406
   \alpha_0
   \alpha_+^{\ast}
   & 0 \\
 0 &
   -0.092902 |\alpha_0|^2
   &
   -0.008406
   \alpha_-
   \alpha_0^{\ast}
   &
   0.092902 |\alpha_0|^2
   & 0 & 0 &
   0.008406\alpha_-
   \alpha_0^{\ast}
   & 0 & 0 \\
 0 & 0 & 0 & 0 & 0 & 0 & 0 & 0
   & 0 \\
 0 & 0 &
   0.008406\alpha_+
   \alpha_0^{\ast}
   & 0 & 0 &
   0.092902 |\alpha_0|^2
   &
   -0.008406
   \alpha_+
   \alpha_0^{\ast}
   &
   -0.092902 |\alpha_0|^2
   & 0 \\
 0 &
   -0.008406
   \alpha_0
   \alpha_-^{\ast}
   &
   -0.000761
   \left(|\alpha_+|^2+
   |\alpha_-|^2
   \right) &
   0.008406\alpha_0
   \alpha_-^{\ast}
   & 0 &
   -0.008406
   \alpha_0
   \alpha_+^{\ast}
   &
   0.000761
  
   \left(|\alpha_+|^2+
   |\alpha_-|^2
   \right) &
   0.008406\alpha_0
   \alpha_+^{\ast}
   & 0 \\
 0 & 0 &
   -0.008406
   \alpha_+
   \alpha_0^{\ast}
   & 0 & 0 &
   -0.092902 |\alpha_0|^2
   &
   0.008406\alpha_+
   \alpha_0^{\ast}
   &
   0.092902 |\alpha_0|^2
   & 0 \\
 0 & 0 & 0 & 0 & 0 & 0 & 0 & 0
   & 0
\end{array}
\right),
\]

\[\fl
A^{(2)}_4=10^{-3}\times
\left(
\begin{array}{lllllllll}

   24.313|\alpha_+|^2
   &
   1.010\alpha_0
   \alpha_+^{\ast}
   &
   20.013\alpha_-
   \alpha_+^{\ast}
   &
   1.010
   \alpha_0
   \alpha_+^{\ast}
   &
   -46.426
   \alpha_-
   \alpha_+^{\ast}
   & 0 &
   20.013\alpha_-
   \alpha_+^{\ast}
   & 0 & 0 \\

   1.010\alpha_+
   \alpha_0^{\ast}
   &
   0.050
    |\alpha_0|^2
   &
   0.905 \alpha_-
   \alpha_0^{\ast}
   &
   0.050
    |\alpha_0|^2
   &
   -2.100
   \alpha_-
   \alpha_0^{\ast}
   & 0 &
   0.905
   \alpha_-
   \alpha_0^{\ast}
   & 0 & 0 \\

   20.013\alpha_+
   \alpha_-^{\ast}
   &
   0.905\alpha_0
   \alpha_-^{\ast}
   &
   16.473 \left(|\alpha_+|^2+
   |\alpha_-|^2
   \right) &
   0.905
   \alpha_0
   \alpha_-^{\ast}
   &
   -38.214\left(|\alpha_+|^2+
   |\alpha_-|^2
   \right) &
   0.905
   \alpha_0
   \alpha_+^{\ast}
   &
   16.473\left(|\alpha_+|^2+
   |\alpha_-|^2
   \right) &
   0.905
   \alpha_0
   \alpha_+^{\ast}
   &
   20.013\alpha_-
   \alpha_+^{\ast}
   \\

   1.010
   \alpha_+
   \alpha_0^{\ast}
   &
   0.050
    |\alpha_0|^2
   &
   0.905
   \alpha_-
   \alpha_0^{\ast}
   &
   0.050    |\alpha_0|^2
   &
   -2.100   \alpha_-
   \alpha_0^{\ast}
   & 0 &
   0.905
   \alpha_-
   \alpha_0^{\ast}
   & 0 & 0 \\

   -46.426
   \alpha_+
   \alpha_-^{\ast}
   &
   -2.100
   \alpha_0
   \alpha_-^{\ast}
   &
   -38.214 \left(|\alpha_+|^2+
   |\alpha_-|^2
   \right) &
   -2.100
   \alpha_0
   \alpha_-^{\ast}
   &
   88.651 \left(|\alpha_+|^2+
   |\alpha_-|^2
   \right) &
   -2.100
   \alpha_0
   \alpha_+^{\ast}
   &
   -38.214
  
   \left(|\alpha_+|^2+
   |\alpha_-|^2
   \right) &
   -2.100
   \alpha_0
   \alpha_+^{\ast}
   &
   -46.426
   \alpha_-
   \alpha_+^{\ast}
   \\
 0 & 0 &
   0.905
   \alpha_+
   \alpha_0^{\ast}
   & 0 &
   -2.100
   \alpha_+
   \alpha_0^{\ast}
   &
   0.050
    |\alpha_0|^2
   &
   0.905   \alpha_+
   \alpha_0^{\ast}
   &
   0.050
    |\alpha_0|^2
   &
   1.100
   \alpha_-
   \alpha_0^{\ast}
   \\

   20.013\alpha_+
   \alpha_-^{\ast}
   &
   0.905
   \alpha_0
   \alpha_-^{\ast}
   &
   16.473\left(|\alpha_+|^2+
   |\alpha_-|^2
   \right) &
   0.905
   \alpha_0
   \alpha_-^{\ast}
   &
   -38.214
  
   \left(|\alpha_+|^2+
   |\alpha_-|^2
   \right) &
   0.905
   \alpha_0
   \alpha_+^{\ast}
   &
   16.473\left(|\alpha_+|^2+
   |\alpha_-|^2
   \right) &
   0.905
   \alpha_0
   \alpha_+^{\ast}
   &
   20.013\alpha_-
   \alpha_+^{\ast}
   \\
 0 & 0 &
   0.905
   \alpha_+
   \alpha_0^{\ast}
   & 0 &
   -2.100
   \alpha_+
   \alpha_0^{\ast}
   &
   0.045
    |\alpha_0|^2
   &
   0.905
   \alpha_+
   \alpha_0^{\ast}
   &
   0.0498
    |\alpha_0|^2
   &
   1.091
   \alpha_-
   \alpha_0^{\ast}
   \\
 0 & 0 &
   20.0126\alpha_+
   \alpha_-^{\ast}
   & 0 &
   -46.4259
   \alpha_+
   \alpha_-^{\ast}
   &
   1.091
   \alpha_0
   \alpha_-^{\ast}
   &
   20.013\alpha_+
   \alpha_-^{\ast}
   &
   1.091   \alpha_0
   \alpha_-^{\ast}
   &
   24.313 |\alpha_-|^2
\end{array}
\right),
\]

\[\fl
A^{(2)}_5=
\left(
\begin{array}{lllllllll}
 \frac{\left(7+4
   \sqrt{3}\right) |\alpha_0|^2}
   {6 \left(9+5
   \sqrt{3}\right)} &
   \frac{\left(2+\sqrt{3}\right) \alpha_-
   \alpha_0^{\ast}}
   {12 \left(9+5
   \sqrt{3}\right)} & 0 &
   \frac{\left(2+\sqrt{3}\right) \alpha_-
   \alpha_0^{\ast}}
   {12 \left(9+5
   \sqrt{3}\right)} & 0 & 0 &
   0 & 0 & 0 \\
 \frac{\left(2+\sqrt{3}\right)
   \alpha_0
   \alpha_-^{\ast}
   }{12 \left(9+5
   \sqrt{3}\right)} &
   \frac{|\alpha_-|^2
   }{216+120 \sqrt{3}} & 0 &
   \frac{|\alpha_-|^2
   }{216+120 \sqrt{3}} & 0 & 0
   & 0 & 0 & 0 \\
 0 & 0 & 0 & 0 & 0 & 0 & 0 & 0
   & 0 \\
 \frac{\left(2+\sqrt{3}\right)
   \alpha_0
   \alpha_-^{\ast}
   }{12 \left(9+5
   \sqrt{3}\right)} &
   \frac{|\alpha_-|^2
   }{216+120 \sqrt{3}} & 0 &
   \frac{|\alpha_-|^2
   }{216+120 \sqrt{3}} & 0 & 0
   & 0 & 0 & 0 \\
 0 & 0 & 0 & 0 & 0 & 0 & 0 & 0
   & 0 \\
 0 & 0 & 0 & 0 & 0 &
   -\frac{\left(-2+\sqrt{3}\right) |\alpha_+|^2}
   {24
   \left(3+\sqrt{3}\right)} &
   0 &
   -\frac{\left(-2+\sqrt{3}\right) |\alpha_+|^2}
   {24
   \left(3+\sqrt{3}\right)} &
   \frac{\alpha_0
   \alpha_+^{\ast}}
   {36+12 \sqrt{3}} \\
 0 & 0 & 0 & 0 & 0 & 0 & 0 & 0
   & 0 \\
 0 & 0 & 0 & 0 & 0 &
   -\frac{\left(-2+\sqrt{3}\right) |\alpha_+|^2}
   {24
   \left(3+\sqrt{3}\right)} &
   0 &
   -\frac{\left(-2+\sqrt{3}\right) |\alpha_+|^2}
   {24
   \left(3+\sqrt{3}\right)} &
   \frac{\alpha_0
   \alpha_+^{\ast}}
   {36+12 \sqrt{3}} \\
 0 & 0 & 0 & 0 & 0 &
   \frac{\alpha_+
   \alpha_0^{\ast}}
   {36+12 \sqrt{3}} & 0 &
   \frac{\alpha_+
   \alpha_0^{\ast}}
   {36+12 \sqrt{3}} &
   \frac{\left(2+\sqrt{3}\right) |\alpha_0|^2}
   {6 \left(3+\sqrt{3}\right)}
\end{array}
\right),
\]

\[\fl
A^{(2)}_6=
\left(
\begin{array}{lllllllll}
 0 & 0 & 0 & 0 & 0 & 0 & 0 & 0
   & 0 \\
 0 & \frac{\left(7+4
   \sqrt{3}\right) |\alpha_-|^2
   }{8 \left(9+5
   \sqrt{3}\right)} & 0 &
   -\frac{\left(7+4
   \sqrt{3}\right) |\alpha_-|^2
   }{8 \left(9+5
   \sqrt{3}\right)} & 0 & 0 &
   0 & 0 & 0 \\
 0 & 0 & 0 & 0 & 0 & 0 & 0 & 0
   & 0 \\
 0 & -\frac{\left(7+4
   \sqrt{3}\right) |\alpha_-|^2
   }{8 \left(9+5
   \sqrt{3}\right)} & 0 &
   \frac{\left(7+4
   \sqrt{3}\right) |\alpha_-|^2
   }{8 \left(9+5
   \sqrt{3}\right)} & 0 & 0 &
   0 & 0 & 0 \\
 0 & 0 & 0 & 0 & 0 & 0 & 0 & 0
   & 0 \\
 0 & 0 & 0 & 0 & 0 &
   \frac{\left(2+\sqrt{3}\right) |\alpha_+|^2}
   {8 \left(3+\sqrt{3}\right)}
   & 0 &
   -\frac{\left(2+\sqrt{3}\right) |\alpha_+|^2}
   {8 \left(3+\sqrt{3}\right)}
   & 0 \\
 0 & 0 & 0 & 0 & 0 & 0 & 0 & 0
   & 0 \\
 0 & 0 & 0 & 0 & 0 &
   -\frac{\left(2+\sqrt{3}\right) |\alpha_+|^2}
   {8 \left(3+\sqrt{3}\right)}
   & 0 &
   \frac{\left(2+\sqrt{3}\right) |\alpha_+|^2}
   {8 \left(3+\sqrt{3}\right)}
   & 0 \\
 0 & 0 & 0 & 0 & 0 & 0 & 0 & 0
   & 0
\end{array}
\right),
\]

\[\fl
A^{(2)}_7=
\left(
\begin{array}{lllllllll}
 0 & 0 & 0 & 0 & 0 & 0 & 0 & 0
   & 0 \\
 0 & \frac{|\alpha_+|^2}
   {48} & \frac{\alpha_0
   \alpha_+^{\ast}}
   {24} & -\frac{|\alpha_+|^2}
   {48} & 0 & \frac{\alpha_-
   \alpha_+^{\ast}}
   {48} & -\frac{\alpha_0
   \alpha_+^{\ast}}
   {24} & -\frac{\alpha_-
   \alpha_+^{\ast}}
   {48} & 0 \\
 0 & \frac{\alpha_+
   \alpha_0^{\ast}}
   {24} & \frac{|\alpha_0|^2}
   {12} & -\frac{\alpha_+
   \alpha_0^{\ast}}
   {24} & 0 & \frac{\alpha_-
   \alpha_0^{\ast}}
   {24} & -\frac{|\alpha_0|^2}
   {12} & -\frac{\alpha_-
   \alpha_0^{\ast}}
   {24} & 0 \\
 0 & -\frac{|\alpha_+|^2}
   {48} & -\frac{\alpha_0
   \alpha_+^{\ast}}
   {24} & \frac{|\alpha_+|^2}
   {48} & 0 &
   -\frac{\alpha_-
   \alpha_+^{\ast}}
   {48} & \frac{\alpha_0
   \alpha_+^{\ast}}
   {24} & \frac{\alpha_-
   \alpha_+^{\ast}}
   {48} & 0 \\
 0 & 0 & 0 & 0 & 0 & 0 & 0 & 0
   & 0 \\
 0 & \frac{\alpha_+
   \alpha_-^{\ast}
   }{48} & \frac{\alpha_0
   \alpha_-^{\ast}
   }{24} & -\frac{\alpha_+
   \alpha_-^{\ast}
   }{48} & 0 &
   \frac{|\alpha_-|^2
   }{48} & -\frac{\alpha_0
   \alpha_-^{\ast}
   }{24} & -\frac{|\alpha_-|^2
   }{48} & 0 \\
 0 & -\frac{\alpha_+
   \alpha_0^{\ast}}
   {24} & -\frac{|\alpha_0|^2}
   {12} & \frac{\alpha_+
   \alpha_0^{\ast}}
   {24} & 0 &
   -\frac{\alpha_-
   \alpha_0^{\ast}}
   {24} & \frac{|\alpha_0|^2}
   {12} & \frac{\alpha_-
   \alpha_0^{\ast}}
   {24} & 0 \\
 0 & -\frac{\alpha_+
   \alpha_-^{\ast}
   }{48} & -\frac{\alpha_0
   \alpha_-^{\ast}
   }{24} & \frac{\alpha_+
   \alpha_-^{\ast}
   }{48} & 0 &
   -\frac{|\alpha_-|^2
   }{48} & \frac{\alpha_0
   \alpha_-^{\ast}
   }{24} & \frac{|\alpha_-|^2
   }{48} & 0 \\
 0 & 0 & 0 & 0 & 0 & 0 & 0 & 0
   & 0
\end{array}
\right),
\]

\[\fl
A^{(2)}_8=
\left(
\begin{array}{lllllllll}
 \frac{|\alpha_-|^2
   }{3} & 0 & 0 & 0 & 0 & 0 &
   0 & 0 & 0 \\
 0 & \frac{|\alpha_+|^2}
   {48} & 0 & \frac{|\alpha_+|^2}
   {48} & 0 &
   -\frac{\alpha_-
   \alpha_+^{\ast}}
   {48} & 0 &
   -\frac{\alpha_-
   \alpha_+^{\ast}}
   {48} & 0 \\
 0 & 0 & 0 & 0 & 0 & 0 & 0 & 0
   & 0 \\
 0 & \frac{|\alpha_+|^2}
   {48} & 0 & \frac{|\alpha_+|^2}
   {48} & 0 &
   -\frac{\alpha_-
   \alpha_+^{\ast}}
   {48} & 0 &
   -\frac{\alpha_-
   \alpha_+^{\ast}}
   {48} & 0 \\
 0 & 0 & 0 & 0 & 0 & 0 & 0 & 0
   & 0 \\
 0 & -\frac{\alpha_+
   \alpha_-^{\ast}
   }{48} & 0 &
   -\frac{\alpha_+
   \alpha_-^{\ast}
   }{48} & 0 &
   \frac{|\alpha_-|^2
   }{48} & 0 &
   \frac{|\alpha_-|^2
   }{48} & 0 \\
 0 & 0 & 0 & 0 & 0 & 0 & 0 & 0
   & 0 \\
 0 & -\frac{\alpha_+
   \alpha_-^{\ast}
   }{48} & 0 &
   -\frac{\alpha_+
   \alpha_-^{\ast}
   }{48} & 0 &
   \frac{|\alpha_-|^2
   }{48} & 0 &
   \frac{|\alpha_-|^2
   }{48} & 0 \\
 0 & 0 & 0 & 0 & 0 & 0 & 0 & 0
   & \frac{|\alpha_+|^2}
   {3}
\end{array}
\right),
\]

\[\fl
\begin{array}{lll}
A^{(2)}_9=10^{-3}\times\\
\left(
\begin{array}{lllllllll}

   30.291 |\alpha_+|^2
   &
   51.354\alpha_0
   \alpha_+^{\ast}
   &
   -0.999 \alpha_-
   \alpha_+^{\ast}
   &
   51.354\alpha_0
   \alpha_+^{\ast}
   &
   42.126 \alpha_-
   \alpha_+^{\ast}
   & 0 &
   -0.999
   \alpha_-
   \alpha_+^{\ast}
   & 0 & 0 \\

   51.354\alpha_+
   \alpha_0^{\ast}
   &
   87.063 |\alpha_0|^2
   &
   -1.694 \alpha_-
   \alpha_0^{\ast}
   &
   87.063 |\alpha_0|^2
   &
   71.418 \alpha_-
   \alpha_0^{\ast}
   & 0 &
   -1.694
   1 \alpha_-
   \alpha_0^{\ast}
   & 0 & 0 \\

   -0.999
   8 \alpha_+
   \alpha_-^{\ast}
   &
   -1.694
   4 \alpha_0
   \alpha_-^{\ast}
   &
   0.033
   \left(|\alpha_+|^2+
   |\alpha_-|^2
   \right) &
   -1.694
   1 \alpha_0
   \alpha_-^{\ast}
   &
   -1.389   \left(|\alpha_+|^2+
   |\alpha_-|^2
   \right) &
   -1.694
    \alpha_0
   \alpha_+^{\ast}
   &
   0.033   
   \left(|\alpha_+|^2+
   |\alpha_-|^2
   \right) &
   -1.694
    \alpha_0
   \alpha_+^{\ast}
   &
   -0.999
   \alpha_-
   \alpha_+^{\ast}
   \\

   51.354\alpha_+
   \alpha_0^{\ast}
   &
   87.063 |\alpha_0|^2
   &
   -1.694
    \alpha_-
   \alpha_0^{\ast}
   &
   87.063 |\alpha_0|^2
   &
   71.418 \alpha_-
   \alpha_0^{\ast}
   & 0 &
   -1.694
    \alpha_-
   \alpha_0^{\ast}
   & 0 & 0 \\

   42.126 \alpha_+
   \alpha_-^{\ast}
   &
   71.418 \alpha_0
   \alpha_-^{\ast}
   &
   -1.389
   
   \left(|\alpha_+|^2+
   |\alpha_-|^2
   \right) &
   71.418 \alpha_0
   \alpha_-^{\ast}
   &
   58.584 \left(|\alpha_+|^2+
   |\alpha_-|^2
   \right) &
   71.418 \alpha_0
   \alpha_+^{\ast}
   &
   -1.389
   
   \left(|\alpha_+|^2+
   |\alpha_-|^2
   \right) &
   71.418 \alpha_0
   \alpha_+^{\ast}
   &
   42.126 \alpha_-
   \alpha_+^{\ast}
   \\
 0 & 0 &
   -1.694
   1 \alpha_+
   \alpha_0^{\ast}
   & 0 &
   71.418 \alpha_+
   \alpha_0^{\ast}
   &
   87.063 |\alpha_0|^2
   &
   -1.694
    \alpha_+
   \alpha_0^{\ast}
   &
   87.063 |\alpha_0|^2
   &
   51.354\alpha_-
   \alpha_0^{\ast}
   \\

   -0.999
   \alpha_+
   \alpha_-^{\ast}
   &
   -1.694
    \alpha_0
   \alpha_-^{\ast}
   &
   0.033
   
   \left(|\alpha_+|^2+
   |\alpha_-|^2
   \right) &
   -1.694
    \alpha_0
   \alpha_-^{\ast}
   &
   -1.389   
   \left(|\alpha_+|^2+
   |\alpha_-|^2
   \right) &
   -1.694
    \alpha_0
   \alpha_+^{\ast}
   &
   0.033
   
   \left(|\alpha_+|^2+
   |\alpha_-|^2
   \right) &
   -1.694
    \alpha_0
   \alpha_+^{\ast}
   &
   -0.999
   \alpha_-
   \alpha_+^{\ast}
   \\
 0 & 0 &
   -1.694
    \alpha_+
   \alpha_0^{\ast}
   & 0 &
   71.418 \alpha_+
   \alpha_0^{\ast}
   &
   87.063 |\alpha_0|^2
   &
   -1.694
    \alpha_+
   \alpha_0^{\ast}
   &
   87.063 |\alpha_0|^2
   &
   51.354\alpha_-
   \alpha_0^{\ast}
   \\
 0 & 0 &
   -0.999
   \alpha_+
   \alpha_-^{\ast}
   & 0 &
   42.126 \alpha_+
   \alpha_-^{\ast}
   &
   51.354\alpha_0
   \alpha_-^{\ast}
   &
   -0.999
   \alpha_+
   \alpha_-^{\ast}
   &
   51.354\alpha_0
   \alpha_-^{\ast}
   &
   30.291 |\alpha_-|^2
\end{array}
\right),
\end{array}
\]

\[\fl
A^{(2)}_{10}=
\left(
\begin{array}{lllllllll}
 0 & 0 & 0 & 0 & 0 & 0 & 0 & 0
   & 0 \\
 0 &
   0.002811
   |\alpha_0|^2
   &
   0.009533 \alpha_-
   \alpha_0^{\ast}
   &
   -0.002811
   |\alpha_0|^2
   & 0 & 0 &
   -0.009533
   \alpha_-
   \alpha_0^{\ast}
   & 0 & 0 \\
 0 &
   0.009533 \alpha_0
   \alpha_-^{\ast}
   &
   0.032322 \left(|\alpha_+|^2+
   |\alpha_-|^2
   \right) &
   -0.009533
   \alpha_0
   \alpha_-^{\ast}
   & 0 &
   0.009533\alpha_0
   \alpha_+^{\ast}
   &
   -0.032322\left(|\alpha_+|^2+
   |\alpha_-|^2
   \right) &
   -0.009533
   \alpha_0
   \alpha_+^{\ast}
   & 0 \\
 0 &
   -0.002811
   |\alpha_0|^2
   &
   -0.009533
   \alpha_-
   \alpha_0^{\ast}
   &
   0.002811
   |\alpha_0|^2
   & 0 & 0 &
   0.009533\alpha_-
   \alpha_0^{\ast}
   & 0 & 0 \\
 0 & 0 & 0 & 0 & 0 & 0 & 0 & 0
   & 0 \\
 0 & 0 &
   0.009533\alpha_+
   \alpha_0^{\ast}
   & 0 & 0 &
   0.002811
   |\alpha_0|^2
   &
   -0.009533
   \alpha_+
   \alpha_0^{\ast}
   &
   -0.002811
   |\alpha_0|^2
   & 0 \\
 0 &
   -0.009533
   \alpha_0
   \alpha_-^{\ast}
   &
   -0.032322\left(|\alpha_+|^2+
   |\alpha_-|^2
   \right) &
   0.009534\alpha_0
   \alpha_-^{\ast}
   & 0 &
   -0.009533
   \alpha_0
   \alpha_+^{\ast}
   &
   0.032322\left(|\alpha_+|^2+
   |\alpha_-|^2
   \right) &
   0.009533\alpha_0
   \alpha_+^{\ast}
   & 0 \\
 0 & 0 &
   -0.009533
   \alpha_+
   \alpha_0^{\ast}
   & 0 & 0 &
   -0.002811
   |\alpha_0|^2
   &
   0.009533\alpha_+
   \alpha_0^{\ast}
   &
   0.002811
   |\alpha_0|^2
   & 0 \\
 0 & 0 & 0 & 0 & 0 & 0 & 0 & 0
   & 0
\end{array}
\right),
\]

\[\fl
A^{(2)}_{11}=10^{-2}\times\left(
\begin{array}{lllllllll}
 0 & 0 & 0 & 0 & 0 & 0 &
   0 & 0 & 0 \\
 0 & 3.442201 \alpha_+ \alpha_+^{\ast} & 4.445611 \alpha_0
   \alpha_+^{\ast} & 3.442201 \alpha_+ \alpha_+^{\ast} & 4.877580
   \alpha_0 \alpha_+^{\ast} & 3.442201 \alpha_- \alpha_+^{\ast} & 4.445611 \alpha_0 \alpha_+^{\ast} & 3.442201 \alpha_- \alpha_+^{\ast} & 0 \\
 0 & 4.445611 \alpha_+ \alpha_0^{\ast} & 5.741519 \alpha_0
   \alpha_0^{\ast} & 4.445611 \alpha_+ \alpha_0^{\ast} & 6.299408 \alpha_0
   \alpha_0^{\ast} & 4.445611 \alpha_- \alpha_0^{\ast} & 5.741519
   \alpha_0 \alpha_0^{\ast} & 4.445611 \alpha_- \alpha_0^{\ast} & 0 \\
 0 & 3.442201 \alpha_+ \alpha_+^{\ast} & 4.445611 \alpha_0
   \alpha_+^{\ast} & 3.442201 \alpha_+ \alpha_+^{\ast} & 4.877580
   \alpha_0 \alpha_+^{\ast} & 3.442201 \alpha_- \alpha_+^{\ast} & 4.445611 \alpha_0 \alpha_+^{\ast} & 3.442201 \alpha_- \alpha_+^{\ast} & 0 \\
 0 & 4.877580 \alpha_+ \alpha_0^{\ast} & 6.299408 \alpha_0
   \alpha_0^{\ast} & 4.877580 \alpha_+ \alpha_0^{\ast} & 6.911506\alpha_0
   \alpha_0^{\ast} & 4.877580 \alpha_- \alpha_0^{\ast} & 6.299408
   \alpha_0 \alpha_0^{\ast} & 4.877580 \alpha_- \alpha_0^{\ast} & 0 \\
 0 & 3.442201 \alpha_+ \alpha_-^{\ast} & 4.445611 \alpha_0
   \alpha_-^{\ast} & 3.442201 \alpha_+ \alpha_-^{\ast} & 4.877580
      \alpha_0 \alpha_-^{\ast} & 3.442201 \alpha_- \alpha_-^{\ast} & 4.445611 \alpha_0 \alpha_-^{\ast} & 3.442201 \alpha_- \alpha_-^{\ast} & 0 \\
 0 & 4.445611 \alpha_+ \alpha_0^{\ast} & 5.741519 \alpha_0
   \alpha_0^{\ast} & 4.445611 \alpha_+ \alpha_0^{\ast} & 6.299408 \alpha_0
   \alpha_0^{\ast} & 4.445611 \alpha_- \alpha_0^{\ast} & 5.741519
   \alpha_0 \alpha_0^{\ast} & 4.445611 \alpha_- \alpha_0^{\ast} & 0 \\
 0 & 3.442201 \alpha_+ \alpha_-^{\ast} & 4.445611 \alpha_0
   \alpha_-^{\ast} & 3.442201 \alpha_+ \alpha_-^{\ast} & 4.877580 
   \alpha_0 \alpha_-^{\ast} & 3.442201 \alpha_- \alpha_-^{\ast} & 4.445611\alpha_0 \alpha_-^{\ast} & 3.442201 \alpha_- \alpha_-^{\ast} & 0 \\
 0 & 0 & 0 & 0 & 0 & 0 &
   0 & 0 & 0
\end{array}
\right)
\]

\[\fl
A^{(2)}_{12}=
\frac{-793451+299804
   \sqrt{7}}{-173551+65377
   \sqrt{7}}\times
\left(
\begin{array}{lllllllll}
 0 & 0 & 0 & 0 & 0 & 0 & 0 & 0
   & 0 \\
 0 &
   \frac{ |\alpha_+|^2}
   {144 } & 0 &
  - \frac{ |\alpha_+|^2}
   {144} & 0 &
  - \frac{\alpha_-
   \alpha_+^{\ast}}
   {144 } & 0 &
   \frac{ \alpha_-
   \alpha_+^{\ast}}
   {144} & 0 \\
 0 & 0 & 0 & 0 & 0 & 0 & 0 & 0
   & 0 \\
 0 & -\frac{ |\alpha_+|^2}
   {144} & 0 &
   \frac{ |\alpha_+|^2}
   {144} & 0 &
   \frac{ \alpha_-
   \alpha_+^{\ast}}
   {144 } & 0 &
   -\frac{ \alpha_-
   \alpha_+^{\ast}}
   {144 } & 0 \\
 0 & 0 & 0 & 0 & 0 & 0 & 0 & 0
   & 0 \\
 0 &- \frac{ \alpha_+
   \alpha_-^{\ast}
   }{144 } & 0 &
   \frac{ \alpha_+
   \alpha_-^{\ast}
   }{144} & 0 &
   \frac{ |\alpha_-|^2
   }{144} & 0 &
   -\frac{ |\alpha_-|^2
   }{144} & 0 \\
 0 & 0 & 0 & 0 & 0 & 0 & 0 & 0
   & 0 \\
 0 &
   \frac{ \alpha_+
   \alpha_-^{\ast}
   }{144} & 0 &
   -\frac{ \alpha_+
   \alpha_-^{\ast}
   }{144} & 0 &
   -\frac{ |\alpha_-|^2
   }{144 } & 0 &
   \frac{ |\alpha_-|^2
   }{144} & 0 \\
 0 & 0 & 0 & 0 & 0 & 0 & 0 & 0
   & 0
\end{array}
\right),
\]

\[\fl
A^{(2)}_{13}=
\left(
\begin{array}{lllllllll}
 0 & 0 & 0 & 0 & 0 & 0 & 0 & 0
   & 0 \\
 0 & \frac{\left(-7+4
   \sqrt{3}\right) |\alpha_-|^2
   }{8 \left(-9+5
   \sqrt{3}\right)} & 0 &
   \frac{\left(7-4
   \sqrt{3}\right) |\alpha_-|^2
   }{8 \left(-9+5
   \sqrt{3}\right)} & 0 & 0 &
   0 & 0 & 0 \\
 0 & 0 & 0 & 0 & 0 & 0 & 0 & 0
   & 0 \\
 0 & \frac{\left(7-4
   \sqrt{3}\right) |\alpha_-|^2
   }{8 \left(-9+5
   \sqrt{3}\right)} & 0 &
   \frac{\left(-7+4
   \sqrt{3}\right) |\alpha_-|^2
   }{8 \left(-9+5
   \sqrt{3}\right)} & 0 & 0 &
   0 & 0 & 0 \\
 0 & 0 & 0 & 0 & 0 & 0 & 0 & 0
   & 0 \\
 0 & 0 & 0 & 0 & 0 &
   \frac{\left(-2+\sqrt{3}\right) |\alpha_+|^2}
   {8
   \left(-3+\sqrt{3}\right)} &
   0 &
   -\frac{\left(-2+\sqrt{3}\right) |\alpha_+|^2}
   {8
   \left(-3+\sqrt{3}\right)} &
   0 \\
 0 & 0 & 0 & 0 & 0 & 0 & 0 & 0
   & 0 \\
 0 & 0 & 0 & 0 & 0 &
   -\frac{\left(-2+\sqrt{3}\right) |\alpha_+|^2}
   {8
   \left(-3+\sqrt{3}\right)} &
   0 &
   \frac{\left(-2+\sqrt{3}\right) |\alpha_+|^2}
   {8
   \left(-3+\sqrt{3}\right)} &
   0 \\
 0 & 0 & 0 & 0 & 0 & 0 & 0 & 0
   & 0
\end{array}
\right),
\]

\[\fl
A^{(2)}_{14}=
\left(
\begin{array}{lllllllll}
 \frac{\left(-7+4
   \sqrt{3}\right) |\alpha_0|^2}
   {6 \left(-9+5
   \sqrt{3}\right)} &
   \frac{\left(-2+\sqrt{3}\right) \alpha_-
   \alpha_0^{\ast}}
   {12 \left(-9+5
   \sqrt{3}\right)} & 0 &
   \frac{\left(-2+\sqrt{3}\right) \alpha_-
   \alpha_0^{\ast}}
   {12 \left(-9+5
   \sqrt{3}\right)} & 0 & 0 &
   0 & 0 & 0 \\

   \frac{\left(-2+\sqrt{3}\right) \alpha_0
   \alpha_-^{\ast}
   }{12 \left(-9+5
   \sqrt{3}\right)} &
   \frac{|\alpha_-|^2
   }{24 \left(9-5
   \sqrt{3}\right)} & 0 &
   \frac{|\alpha_-|^2
   }{24 \left(9-5
   \sqrt{3}\right)} & 0 & 0 &
   0 & 0 & 0 \\
 0 & 0 & 0 & 0 & 0 & 0 & 0 & 0
   & 0 \\

   \frac{\left(-2+\sqrt{3}\right) \alpha_0
   \alpha_-^{\ast}
   }{12 \left(-9+5
   \sqrt{3}\right)} &
   \frac{|\alpha_-|^2
   }{24 \left(9-5
   \sqrt{3}\right)} & 0 &
   \frac{|\alpha_-|^2
   }{24 \left(9-5
   \sqrt{3}\right)} & 0 & 0 &
   0 & 0 & 0 \\
 0 & 0 & 0 & 0 & 0 & 0 & 0 & 0
   & 0 \\
 0 & 0 & 0 & 0 & 0 &
   -\frac{\left(2+\sqrt{3}\right) |\alpha_+|^2}
   {24
   \left(-3+\sqrt{3}\right)} &
   0 &
   -\frac{\left(2+\sqrt{3}\right) |\alpha_+|^2}
   {24
   \left(-3+\sqrt{3}\right)} &
   \frac{\alpha_0
   \alpha_+^{\ast}}
   {36-12 \sqrt{3}} \\
 0 & 0 & 0 & 0 & 0 & 0 & 0 & 0
   & 0 \\
 0 & 0 & 0 & 0 & 0 &
   -\frac{\left(2+\sqrt{3}\right) |\alpha_+|^2}
   {24
   \left(-3+\sqrt{3}\right)} &
   0 &
   -\frac{\left(2+\sqrt{3}\right) |\alpha_+|^2}
   {24
   \left(-3+\sqrt{3}\right)} &
   \frac{\alpha_0
   \alpha_+^{\ast}}
   {36-12 \sqrt{3}} \\
 0 & 0 & 0 & 0 & 0 &
   \frac{\alpha_+
   \alpha_0^{\ast}}
   {36-12 \sqrt{3}} & 0 &
   \frac{\alpha_+
   \alpha_0^{\ast}}
   {36-12 \sqrt{3}} &
   \frac{\left(-2+\sqrt{3}\right) |\alpha_0|^2}
   {6
   \left(-3+\sqrt{3}\right)}
\end{array}
\right),
\]

\[\fl
A^{(2)}_{15}=
\left(
\begin{array}{lllllllll}
 0 & 0 & 0 & 0 & 0 & 0 & 0 & 0
   & 0 \\
 0 &
   0.001509
   |\alpha_0|^2
   &
   0.009839\alpha_-
   \alpha_0^{\ast}
   &
   -0.001509
   |\alpha_0|^2
   & 0 & 0 &
   -0.009839\alpha_-
   \alpha_0^{\ast}
   & 0 & 0 \\
 0 &
   0.009839\alpha_0
   \alpha_-^{\ast}
   &
   0.064140 \left(|\alpha_+|^2+
   |\alpha_-|^2
   \right) &
   -0.009839
   \alpha_0
   \alpha_-^{\ast}
   & 0 &
   0.009839\alpha_0
   \alpha_+^{\ast}
   &
   -0.064140\left(|\alpha_+|^2+
   |\alpha_-|^2
   \right) &
   -0.009839
   \alpha_0
   \alpha_+^{\ast}
   & 0 \\
 0 &
   -0.001509
   |\alpha_0|^2
   &
   -0.009840
   \alpha_-
   \alpha_0^{\ast}
   &
   0.001509
   |\alpha_0|^2
   & 0 & 0 &
   0.009839 \alpha_-
   \alpha_0^{\ast}
   & 0 & 0 \\
 0 & 0 & 0 & 0 & 0 & 0 & 0 & 0
   & 0 \\
 0 & 0 &
   0.009839\alpha_+
   \alpha_0^{\ast}
   & 0 & 0 &
   0.001509   |\alpha_0|^2
   &
   -0.009839\alpha_+
   \alpha_0^{\ast}
   &
   -0.001509
   |\alpha_0|^2
   & 0 \\
 0 &
   -0.009839\alpha_0
   \alpha_-^{\ast}
   &
   -0.064140\left(|\alpha_+|^2+
   |\alpha_-|^2
   \right) &
   0.009840\alpha_0
   \alpha_-^{\ast}
   & 0 &
   -0.009839\alpha_0
   \alpha_+^{\ast}
   &
   0.064140 \left(|\alpha_+|^2+
   |\alpha_-|^2
   \right) &
   0.009839 \alpha_0
   \alpha_+^{\ast}
   & 0 \\
 0 & 0 &
   -0.009839
   \alpha_+
   \alpha_0^{\ast}
   & 0 & 0 &
   -0.001509
   |\alpha_0|^2
   &
   0.009839 \alpha_+
   \alpha_0^{\ast}
   &
   0.001509
   |\alpha_0|^2
   & 0 \\
 0 & 0 & 0 & 0 & 0 & 0 & 0 & 0
   & 0
\end{array}
\right),
\]

\[\fl
A^{(2)}_{16}= 10^{-3}\times
\left(
\begin{array}{lllllllll}

   0.952
   |\alpha_+|^2
   &
   3.102
   \alpha_0
   \alpha_+^{\ast}
   &
   8.764\alpha_-
   \alpha_+^{\ast}
   &
   3.102   \alpha_0
   \alpha_+^{\ast}
   &
   4.300\alpha_-
   \alpha_+^{\ast}
   & 0 &
   8.764\alpha_-
   \alpha_+^{\ast}
   & 0 & 0 \\

   3.102
   \alpha_+
   \alpha_0^{\ast}
   &
   10.110|\alpha_0|^2
   &
   28.566\alpha_-
   \alpha_0^{\ast}
   &
   10.110 |\alpha_0|^2
   &
   14.016\alpha_-
   \alpha_0^{\ast}
   & 0 &
   28.566\alpha_-
   \alpha_0^{\ast}
   & 0 & 0 \\

   8.764\alpha_+
   \alpha_-^{\ast}
   &
   28.566\alpha_0
   \alpha_-^{\ast}
   &
   80.716 \left(|\alpha_+|^2+
   |\alpha_-|^2
   \right) &
   28.566 \alpha_0
   \alpha_-^{\ast}
   &
   39.604 \left(|\alpha_+|^2+
   |\alpha_-|^2
   \right) &
   28.566 \alpha_0
   \alpha_+^{\ast}
   &
   80.716 \left(|\alpha_+|^2+
   |\alpha_-|^2
   \right) &
   28.566 \alpha_0
   \alpha_+^{\ast}
   &
   8.764\alpha_-
   \alpha_+^{\ast}
   \\

   3.102   \alpha_+
   \alpha_0^{\ast}
   &
   10.110 |\alpha_0|^2
   &
   28.566 \alpha_-
   \alpha_0^{\ast}
   &
   10.110|\alpha_0|^2
   &
   14.016\alpha_-
   \alpha_0^{\ast}
   & 0 &
   28.566\alpha_-
   \alpha_0^{\ast}
   & 0 & 0 \\

   4.300\alpha_+
   \alpha_-^{\ast}
   &
   14.016\alpha_0
   \alpha_-^{\ast}
   &
   39.604 \left(|\alpha_+|^2+
   |\alpha_-|^2
   \right) &
   14.016\alpha_0
   \alpha_-^{\ast}
   &
   19.431\left(|\alpha_+|^2+
   |\alpha_-|^2
   \right) &
   14.016\alpha_0
   \alpha_+^{\ast}
   &
   39.604 \left(|\alpha_+|^2+
   |\alpha_-|^2
   \right) &
   14.016\alpha_0
   \alpha_+^{\ast}
   &
   4.300\alpha_-
   \alpha_+^{\ast}
   \\
 0 & 0 &
   28.566 \alpha_+
   \alpha_0^{\ast}
   & 0 &
   14.016\alpha_+
   \alpha_0^{\ast}
   &
   10.110|\alpha_0|^2
   &
   28.566\alpha_+
   \alpha_0^{\ast}
   &
   10.110 |\alpha_0|^2
   &
   3.102
   \alpha_-
   \alpha_0^{\ast}
   \\

   8.764\alpha_+
   \alpha_-^{\ast}
   &
   28.566\alpha_0
   \alpha_-^{\ast}
   &
   80.716 \left(|\alpha_+|^2+
   |\alpha_-|^2
   \right) &
   28.566\alpha_0
   \alpha_-^{\ast}
   &
   39.604 \left(|\alpha_+|^2+
   |\alpha_-|^2
   \right) &
   28.566\alpha_0
   \alpha_+^{\ast}
   &
   80.716 \left(|\alpha_+|^2+
   |\alpha_-|^2
   \right) &
   28.566\alpha_0
   \alpha_+^{\ast}
   &
   8.764\alpha_-
   \alpha_+^{\ast}
   \\
 0 & 0 &
   28.566 \alpha_+
   \alpha_0^{\ast}
   & 0 &
   14.016\alpha_+
   \alpha_0^{\ast}
   &
   10.110 |\alpha_0|^2
   &
   28.566\alpha_+
   \alpha_0^{\ast}
   &
   10.110|\alpha_0|^2
   &
   3.102
   \alpha_-
   \alpha_0^{\ast}
   \\
 0 & 0 &
   8.764\alpha_+
   \alpha_-^{\ast}
   & 0 &
   4.300\alpha_+
   \alpha_-^{\ast}
   &
   3.102
   \alpha_0
   \alpha_-^{\ast}
   &
   8.764\alpha_+
   \alpha_-^{\ast}
   &
   3.102   \alpha_0
   \alpha_-^{\ast}
   &
   0.952
   |\alpha_-|^2
\end{array}
\right).
\]
}}


\normalsize
\section*{References}
\begin{thebibliography}{10}

\bibitem{Wen}
X.-G. Wen, \emph{Quantum Field Theory of Many-Body Systems},
Oxford University Press, Oxford (2004).
\bibitem{Auerbach:94} A. Auerbach, \emph{Interacting Electrons and Quantum Magnetism}, Springer-Verlag, New York (1994). 
\bibitem{Vidal:03} G. Vidal, Phys. Rev. Lett. {\bf 91}, 147902 (2003).
\bibitem{Cirac:06} F. Verstraete and J.I. Cirac, Phys. Rev. B {\bf 73}, 094423 (2006).
\bibitem{Osborn:06} T. Osborne, Phys. Rev. Lett. {\bf 97}, 157202 (2006).
\bibitem{Cirac:04}
F. Verstraete, M.A. Martin-Delgado, and J.I. Cirac, Phys. Rev. Lett. {\bf 92}, 087201 (2004).
\bibitem{Lloyd:96}
S. Lloyd, Science {\bf 273}, 1073 (1996).
\bibitem{Nielsen}
M. A. Nielsen, M. J. Bremner, J. L. Dodd, A. M. Childs, and C. M. Dawson, Phys. Rev. A {\bf 66}, 022317 (2002).
%\bibitem{Knill:02}
%R. Somma, G. Ortiz, J.I. Gubernatis, E. Knill, and R. Laflamme, Phys. Rev. A, {\bf 65}, 042323 (2002).
\bibitem{Brown:06} K.R. Brown, R.J. Clark, and I.L, Chuang, Phys. Rev. Lett. {\bf 97}, 050504 (2006). 
\bibitem{Duan:03} Duan, L.M.~, Demler, E., \& Lukin, M.D., 
  Phys. Rev. Lett. {\bf 91}, 090402 (2003).
\bibitem{Buechler:04} B\"uchler, H.P., Hermele, M., Huber, S.D.,
  Fisher, M.P.A., \& Zoller, P., Phys. Rev. Lett. {\bf 95} , 040402 (2005).
    \bibitem{Cirac:04a} J.J. Garc\'{\i}ia-Ripoll, M.A. Mart\'{\i}n-Delgado, and J.I. Cirac, Phys. Rev. Lett. {\bf 93}, 250405 (2004).
\bibitem{Demler:03} A. Imambekov, M. Lukin, and E. Demler, Phys. Rev. A {\bf 68}, 063602 (2003).
\bibitem{Julienne:06} R. Ciurylo, E. Tiesinga, and P.S. Julienne, Phys. Rev. A {\bf 74}, 022710 (2006).
\bibitem{Deutsch:03}  R. Stock, I.H. Deutsch, and E.L. Bolda, Phys. Rev. Lett. {\bf 91}, 183201 (2003).
\bibitem{Micheli:06} A. Micheli, G.K. Brennen, and P. Zoller, Nature Physics {\bf 2}, 341 (2006). 
\bibitem{review} See Eur. Phys. J. D \emph{Special issue:  Ultracold Polar Molecules: Formation and Collisions},  {\bf 31} (2004).
 \bibitem{DeMille:05} J.M. Sage, S. Sainis, T. Bergeman, and D. DeMille, 
Phys. Rev. Lett. {\bf 94} 203001 (2005).
\bibitem{Haldane:88} F.D.M. Haldane, \emph{Two-Dimensional Strongly Correlated Electron Systems}, edited by Z.Z. Gan and Z.B. Su, Gordon and Breach (1988).
\bibitem{Rad:64}  H.F. Radford, Phys. Rev. {\bf 136} 1571 (1964) 
\bibitem{Ryzlewicz:82} Ch. Ryzlewicz, H.-U. Sch\"utze-Pahlmann, J. Hoeft, and T. T\"orring, Chemical Physics, {\bf 71}, 389 (1982).
\bibitem{Moller:82}  K. M\"oller, H.U. Sch\"utze-Pahlmann, J. Hoeft, and T. T\"orring, Chem. Phys. {\bf 68}, 399 (1982);  W.E. Ernst, S. Kindt, K.P.R. Nair, and T. T\"orring, Phys. Rev. A {\bf 29}, 1158 (1984).  See also \href{http://physics.nist.gov/PhysRefData/MolSpec/Diatomic}
{http://physics.nist.gov/PhysRefData/MolSpec/Diatomic}.
\bibitem{Brennen} G.K. Brennen, I.H. Deutsch, and C.J. Williams, Phys. Rev. A {\bf 65}, 022313 (2002).
\bibitem{Schollwock:96} U. Schollw\"ock, Th. Jolicoeur, and T. Garel, Phys. Rev. B {\bf 53}, 3304 (1996).
\bibitem{Schmitt:98} A. Schmitt, K.-H. M\"utter, M. Karbach, Y. Yu, and G. M\"uller, Phys. Rev. B {\bf 58}, 5498 (1998).
\bibitem{AKLT:87}  I. Affleck, T. Kennedy, E.H. Lieb, and H. Tasaki, Commun. Math. Phys. {\bf 155}, 477 (1988).
\bibitem{Milburn} J.P. Barjaktarevic, R.H. McKenzie, J. Links, and G.J. Milburn, Phys. Rev. Lett. {\bf 95} 230501 (2005).
\bibitem{Fath:93} G. F\'ath and J. S\'olyom, J. Phys. Condens. Matter {\bf 5}, 8983 (1993).
\bibitem{future} Extensions to include many body terms in the effective Hamiltonian are considered in  H.P. B\"uchler, G.K. Brennen, A.M. Micheli, and P. Zoller  (in preparation).
\bibitem{Vidal:04} G. Vidal, Phys. Rev. Lett. {\bf 93}, 040502 (2004).
%\bibitem{Kolezhuk:02} A.K. Kolezhuk and U. Schollw\"ock {\bf 65}, 100401 (2002).
\bibitem{Lai:74} C.K. Lai, J. Math. Phys. {\bf 15}, 1675 (1974); B. Sutherland, Phys. Rev. B {\bf 12}, 3795 (1975).
\bibitem{Grondalski:99}  J. Grondalski, P.M. Alsing, and I.H. Deutsch, Optics Express {\bf 5}, 249 (1999).
\bibitem{Altman:04}  E. Altman, E. Demler, and M.D. Lukin, Phys. Rev. A {\bf 70}, 013603 (2004).
\bibitem{Bloch:05} I. Bloch, Nature Physics {\bf 1}, 23 (2005).
%\bibitem{Duan:06} L.-M. Duan, Phys. Rev. Lett. {\bf 96}, 103201 (2006).
\bibitem{DeMille:02}  D. DeMille, Phys. Rev. Lett. {\bf 88}, 067901 (2002).
\bibitem{Buchler:06} H.P. B\"uchler, E. Demler, M. Lukin, A. Micheli, N. Prokof'ev, G. Pupillo, and P. Zoller, cond-mat/0607294.
\bibitem{Wen:03}  X.-G. Wen, Phys. Rev. B {\bf 68}, 115413 (2003). 

\end{thebibliography}
\end{document}